\documentclass[aps,floatfix,showpacs]{revtex4-1}
\usepackage{graphicx}
\usepackage{dcolumn}
\usepackage{bm}
\usepackage{amsmath}
\usepackage{epstopdf}
\usepackage{hyperref}

\newcommand{\be}{\begin{equation}}
\newcommand{\ee}{\end{equation}}

\begin{document}

\title{Dynamical and statistical phenomena of circulation and heat transfer in periodically forced rotating turbulent Rayleigh-B\'enard convection}

\author{Sebastian Sterl$^{1,2}$\footnote{$^*$These authors contributed equally to this study}\footnote{Present address: NewClimate Institute for Climate Policy and Global Sustainability gGmbH, Am Hof 20-26, 50667 Cologne, Germany}, Hui-Min Li$^{1a}$, Jin-Qiang Zhong$^1$\footnote{Corresponding author, email: \href{mailto:jinqiang@tongji.edu.cn}{jinqiang@tongji.edu.cn}}}

 
\affiliation{$^1$Fluid Lab, Shanghai Key Laboratory of Special Artificial Microstructure Materials and Technology and
School of Physics Science and Engineering, Tongji University, Shanghai 200092, China}
\affiliation{$^2$Physics of Fluids Group, Faculty of Science and Technology, University of Twente, PO Box 217,
7500 AE Enschede, The Netherlands}

\date{\today}
 
\begin{abstract}
In this paper, we present results from an experimental study into turbulent Rayleigh-B\'enard convection forced externally by periodically modulated unidirectional rotation rates. We find that the azimuthal rotation velocity $\dot{\theta}(t)$ and thermal amplitude $\delta(t)$ of the large-scale circulation (LSC) are modulated by the forcing, exhibiting a variety of dynamics including increasing phase delays and a resonant peak in the amplitude of $\dot{\theta}(t)$. 
We also focus on the influence of modulated rotation rates on the frequency of occurrence $\eta$ of stochastic cessation/reorientation events, and on the interplay between such events and the periodically modulated response of $\dot{\theta}(t)$. Here we identify a mechanism by which $\eta$ can be amplified by the modulated response and these normally stochastic events can occur with high regularity. 
We provide a modeling framework that explains the observed amplitude and phase responses, and extend this approach to make predictions for the occurrence of cessation events and the probability distributions of $\dot{\theta}(t)$ and $\delta(t)$ during different phases of a modulation cycle, based on an adiabatic approach that treats each phase separately. 
Lastly, we show that such periodic forcing has consequences beyond influencing LSC dynamics, by investigating how it can modify the heat transport even under conditions where the Ekman pumping effect is predominant and strong enhancement of heat transport occurs. We identify phase and amplitude responses of the heat transport, and show how increased modulations influence the average Nusselt number.
\end{abstract}


\maketitle

\section{Introduction}

Thermal convection is ubiquitous and underlies many important features of natural flows. It occurs on large scales in the atmosphere and oceans and has short-term as well as long-term impacts on weather and climate \cite{Va06, MS99}. It takes place on small scales in biochemical systems and drives the polymerase chain reaction of DNA replication \cite{KUB02, BGL03}. It also plays an important role in many industrial processes, where both the enhancement and inhibition of heat transport may have significant applications.

The quintessential laboratory experiment to investigate thermal convection is the extensively-studied Rayleigh-B\'enard convection (RBC) system, in which a fluid inside a closed container is heated from the bottom and cooled from the top \cite{AGL09, LX10, CS12}. In such a closed system, a convective large-scale circulation (LSC) in the fluid column can appear at sufficiently high temperature differences between the top and bottom plate, presenting a relatively well-defined flow pattern in a background of highly turbulent fluid. This LSC is manifested as a convection roll whose size is comparable to that of the RBC cell. In many studies, the LSC has been modeled as a circulation in a vertical plane, carrying hot fluid up near the bottom side of the sample and cold fluid down on the top side \cite[for examples, see][]{KH81, SBN02, BNA05, XZZCX09}.

In many astrophysical and geophysical systems, thermal convection is strongly influenced by the background rotations \cite{SH00, HBL00, MS99}. In the recent past, a substantive body of research work has been devoted to exploring the dynamical behavior of an LSC in a rotating RBC setup, and its role in overall heat transport. This has involved elaborate studies on the azimuthal rotations of the LSC flow and its thermal strength \cite{HKO02, BA06a, BA06b, BA07, XZX06, KCG08, ZA10, WA11}, the structure of thermal and momentum boundary layers under external rotation \cite{SCL10, KSOSHC11, KCH13}, and the influence of rotation on the statistical responses of LSC orientation and strength \cite{BA06a, AAG11, AAG12}. 

Motivated by its broad geophysical relevance, in this paper we extend on the previous research works and consider the influence of time-varying rotations on turbulent RBC. In the geophysical context, the adjustment of a fluid column to a change in its rotation is an important process in oceanography, primarily in studies on wind-stress-driven flows in the upper oceans \cite{AG75, LH84, FUB02} and their influence on large-scale phenomena such as El Ni\~no \cite{WM74}. Since such geophysical flows are often influenced by thermal convection, their responses could potentially be better understood by studying the fluid dynamics in turbulent RBC with time-varying rotations.  

In the astrophysical context, many celestial bodies themselves do not have a constant rotation rate; the gravitational interaction of a planet with its satellites and other neighbors, for example, can force a periodic variation of its rotation rate (libration), thereby potentially influencing large-scale thermally driven systems on its surface or in its interior \cite{NHWBA09, NCLCA10, SCMB10, ZCL11, CVH14}. An example of a strongly librating body with a liquid interior is planet Mercury \cite{MPJSH07}. The accelerations generated by the time-varying spinning rate may modify the convective flow structures in Mercury's molten core and could have considerable influence on its global magnetic field. A review article \cite{Ol13} summarizes the existing and ongoing laboratory investigations of planetary core dynamics, and discusses the effects that libration has on the flow structures in rotating convection systems.

From the point of view of the fundamental interest in studying turbulent systems, turbulent flows are often subject to various types of periodic modulation. Examples include the Earth's tidal ocean currents, the atmospheric flows periodically forced by solar radiation \cite{JX08}, and the pulsatile blood flow through arteries \cite{Ku97}. If the modulation is slow, i.e.\ when the modulation period is larger than the dominant internal time scales of the flow, the flows can adjust ``adiabatically" to the different states under various rotation rates. In a turbulent RB system rotating at constant rate, the potential presence of a large-scale circulation in conjunction with a turbulent background makes for a situation in which the dynamics of the LSC can be well described by stochastic equations for the diffusive LSC orientation and strength \cite{SBN02, Be05, BA07, BA08a, AAG11, AAG12}. How are the dynamical and statistical properties of the LSC influenced by external forcing (such as from time-dependent rotation)? Under the adiabatic approximation, are the existing low-dimensional models still capable to predict the dynamical behavior of the LSC flow that is subjected to modulated rotations? These are the intriguing problems we will address in this work.     

In this study, we investigate the effects of time-varying (unidirectional) rotation rates on the dynamical as well as statistical behavior of the LSC in a turbulent background under the influence of periodically-modulated rotations. While there exists a body of previous research works, both experimental and numerical, on RB convection with time-dependent rotation, such works have mostly focused on non-turbulent states \cite{NSD91, TBA02, RR02, RLM08, RLM09}. Recently, however, DNS studies \cite{GK14a, GK14b} and an exploratory experimental study \cite{NBS10} have shown potentially significant effects of modulated rotation on heat transport in turbulent RB convection.

To our knowledge, our study is the first full experimental study into the effects of modulated rotation rates on the dynamical and statistical LSC behaviour in turbulent RB convection. A selection of initial results from this study has recently been published in \cite{ZSL15}. The present paper goes into more depth on the methodology of the results and greatly expands upon the previous short paper by providing complete results on the experimental and theoretical investigation of the dynamical and statistical responses of various LSC parameters. 
We describe a wide range of experimentally observed phenomena, ranging from the amplitude and phase responses of LSC strength and orientation, to a possible resonant interplay between modulated flow responses on the one hand, and normally stochastic cessation and reorientation events on the other. We provide extensions of previous modeling approaches to explain the various dynamical and statistical phenomena observed in a consistent manner throughout. Lastly, we move to a different parameter range to provide an initial investigation of the dynamical and statistical response of heat transport in turbulent RB convection with modulated rotation in the absence of an LSC, to show how the effects of modulation go beyond influencing large-scale flow structures.

This paper is structured as follows. Section~II provides details on the experimental setup and methods. Section~III explains the experimental results pertaining to the responses of LSC strength and azimuthal orientation/velocity under the influence of modulated rotation. (These results have been discussed much more briefly in \cite{ZSL15}). Section~IV provides the modeling approach used to explain the results from Section~III. Section~V provides experimental results pertaining to the statistical responses of LSC strength and velocity undergoing modulated rotations, and the role of stochastic cessation events therein. Section~VI extends the modeling approach from Section~IV to explain the observed statistical phenomena in a consistent manner. Section~VII details the experimental results from an exploratory investigation of the influence of modulated rotations on heat transfer in turbulent RB convection. Lastly, conclusions and recommendations for future research are given in Section~VIII.

\section{Experimental apparatus and methods}

\begin{figure}[t!]
\centering
\includegraphics[width=0.4\textwidth]{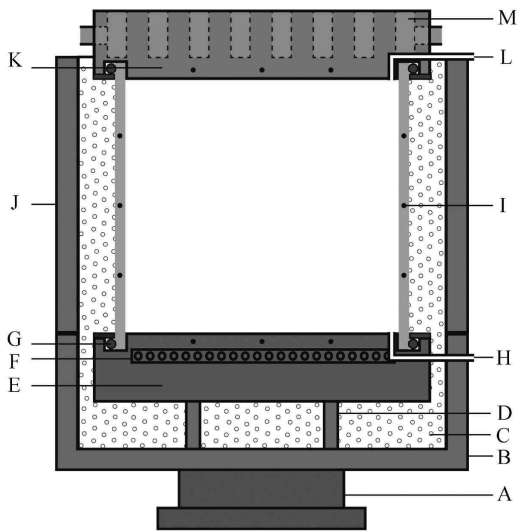}
\caption{A schematic of the experimental setup, as described in the text. (A) rotary table; (B) bottom thermal shield; (C) foam sheet; (D) supporting ring; (E) bottom plate; (F) main heater in bottom plate; (G) O-ring; (H) capillary fluid inlet; (I) sidewall cylinder; (J) thermal side shield; (K) top plate; (L) capillary fluid outlet; (M) double-spiral water-cooling channel.}
\label{fig:Figure_1}
\end{figure}

A schematic diagram of the experimental apparatus used for this study is shown in Fig.~\ref{fig:Figure_1}. A rotary table (A) rested securely on the laboratory floor. Its rotating axis was adjusted accurately to be parallel with gravity. Supported on A was the bottom thermal shield (B) of the convection system. Two heaters made of resistance wires were separately contained inside the bottom and the periphery of shield B, respectively (not shown in the Figure). Thermistors were installed in various locations inside B. During experiments, the input power to the two heaters was controlled such that the temperature in the whole volume of B remained the same as the bottom plate (E) temperature ($T_b$), with an accuracy better than 0.01~K. By virtue of this temperature regulation method, the heat loss through the bottom plate E to the shield B was reduced such as to become essentially negligible.

On top of the shield B, the bottom plate E of the convection cell was supported on a ring D made of bakelite. Bakelite has a high rigidity and a tensile strength comparable to steel, but a much lower thermal conductivity. Thus, the bakelite ring served as a rigid supporting base of the cell with desirable thermal insulation. The bottom plate~E was made of oxygen-free copper (OFHC, type TU1). It had a thickness of 35.0~mm and a diameter of 285.7~mm. Its central area of 242.5~mm diameter was covered uniformly by parallel straight grooves connected by semicircles at their ends. A main heater (F) made of resistance wire with a diameter of 1.0~mm was embedded and epoxied into the grooves. Seven thermistors were installed in the bottom plate, one at the center and the other six equally spaced on a circle of 210.0~mm diameter. Temperature inhomogeneity on the plate, as measured by these thermistors, was within one or two per cent of ${\Delta}T$ the temperature difference between the top and bottom plate, during experiments.

A central section of the plate E, 242.5~mm in diameter, was closely fitted into the sidewall cylinder (I). On one point of the side of the central section in E, there was a capillary (H) of 1.0~mm in diameter through which the fluid entered the system. The cylindrical sidewall was made of Plexiglas and had a wall thickness of 4.0~mm. A nitrile-butadiene rubber O-ring (G) sealed the fluid from outside the sidewall. A similar construction was used to terminate the sidewall near the top plate (K).

The top plate (K) was made of OFHC copper, similar to the bottom plate in its dimensions. It had a double-spiral water-cooling channel (M) machined directly into it from the top. A constant temperature in K ($T_t$) was maintained by circulating coolant in channel M driven by a refrigerating circulator (PolyScience PP30R). The circulation flow speed of the coolant was further enhanced through a programmable fluid pump. A capillary fluid outlet (L) and seven thermistors were installed in K at positions similar to those in the bottom plate. Temperature inhomogeneities in the top plate were about twice larger than in the bottom plate. Thermal protection to the side of the cell was provided by a thermal side-shield (J) made of aluminium. Its temperature, controlled by a second coolant-circulation system, was maintained at the same value as the mean fluid temperature in the cell, to an accuracy of 0.01~K. To reduce heat lost through air convection in the vicinity of the cell, the space outside the cell but inside the shields (B and L) was filled with low-density foam sheet (C). During the experiment, the two sets of coolant circulating circuits as well as all the electrical wires were brought into the convection system through a rotary feed-through built into the table (A).

The experiments pertaining to LSC responses (sections III and V) were performed with a temperature difference $\Delta T = T_b - T_t = 16.00$~K, giving a Rayleigh number $Ra = \alpha g \Delta T L^3 / \kappa\nu  = 8.24\times 10^9$ ($g$~is the gravitational acceleration; $\alpha$, $\nu$  and $\kappa$~are the thermal expansion coefficient, the kinematic viscosity, and thermal diffusivity of water, respectively; $L$ the sample height), with the Prandtl number constant at $Pr = \nu/\kappa = 4.38$. The experiments on heat transfer (section VII) were performed with a four times smaller value $\Delta T = 4.00$~K, yielding $Ra = 2.06 \times 10^9$.

The sample had a diameter $D = 240.0$~mm and a height $L = 240.0$~mm, yielding an aspect ratio $\Gamma = 1.00$. Three rows of thermistors (eight on each row), equally spaced azimuthally and lined up in vertical columns at heights $L/4$, $L/2$ and $3L/4$, were installed into the sidewalls. During experiments, we measured the temperature of each thermistor $T_i$, and fit the function $T_i = T_0 + \delta \cos{(i\pi/4 - \theta)}$, $i = 1, ...8$, to the eight temperatures in each row. Following this method, as used before in \cite{BA06a, ZA10}, the thermal amplitude~$\delta(t)$ of a large-scale circulation (LSC), and the azimuthal orientation~$\theta(t)$ of its circulating plane (as seen from the rotating frame of reference), could be determined. (The results shown in this paper are measurements from the middle-row thermistors unless otherwise noted. However, results from the top and bottom thermistor row were always used for consistency checks with the middle row.) 

When working in a modulation mode, the rotating velocity of the sample was varied periodically according to $\Omega(t) = \Omega_0[1 + \beta\cos{(\omega t)}]$, with $\beta < 1$ to ensure unidirectional modulation. More information on the experimental protocol to obtain such rotation rates is given in Appendix~A.

The heat transfer in the sample can be expressed by the dimensionless Nusselt number $Nu$, which is the ratio of the total heat transfer to the purely conductive heat transfer that would occur in the absence of any convection (i.e. below the convective instability threshold). Hence it is given by $Nu = QL/(\Delta T \lambda)$, where $Q$ is the vertical heat flux and $\lambda$ is the thermal conductivity of the fluid. In this experimental setup, $Q$ is determined by the input power to the heater F in the bottom plate (with appropriate corrections~\cite{ZA10}), the value of which is rigorously controlled through digital feedback on the basis of the requirement that the bottom plate temperature $T_b$ remain constant throughout an experiment.

\section{Experimental results: periodic response of the LSC azimuthal velocity and amplitude}

In this section, we discuss our experimental results pertaining to the influence of modulated rotation rates on the dynamics of the large-scale circulation. We firstly focus on the azimuthal LSC velocity $\dot{\theta}$ and the thermal LSC strength $\delta$ under constant-rotation conditions. Afterwards, we present our results on modulated-rotation and compare them to the constant-rotation case.

\subsection{Results for constant rotation}

\begin{figure}[t!]
\centering
\includegraphics[width=0.8\textwidth]{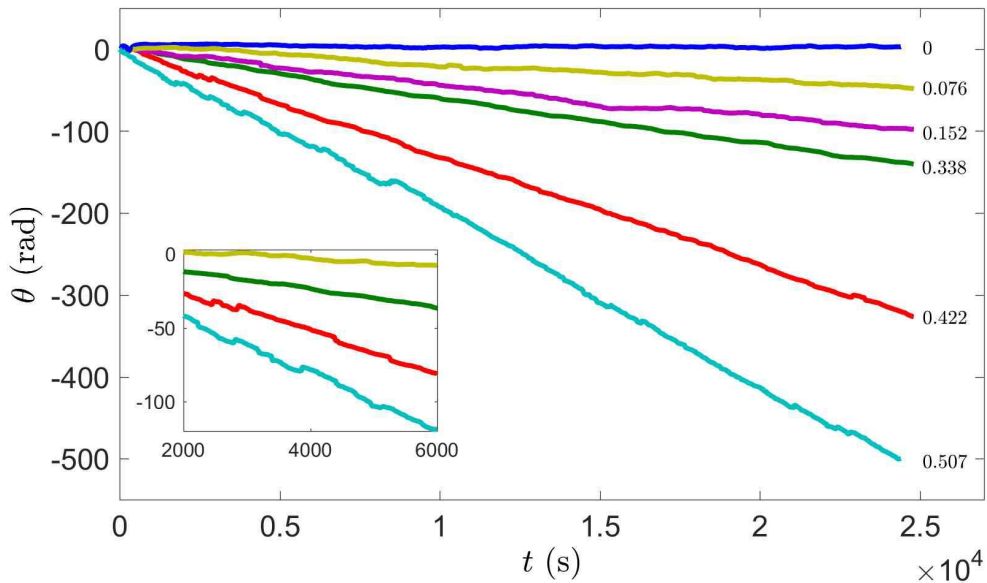}
\caption{The LSC orientation $\theta$ (always retrograde as seen from the rotating frame) with respect to time, obtained from middle-thermistor data. The values next to the curves indicate the value of $1/Ro$. We have arbitrarily defined $\theta(0) = 0$ for each curve. (inset) Close-up for $1/Ro = 0.076$, 0.338, 0.422 and 0.507 (from top to bottom) on a shorter timescale, showing how the linear trend is significantly affected by diffusive motions.}
\label{fig:Figure_2}
\end{figure}

In Fig.~\ref{fig:Figure_2}, we plot the orientation $\theta(t)$ for a number of experiments. The values next to the curves indicate the corresponding value of $1/Ro$. The curve corresponding to $1/Ro = 0$ consists solely of fluctuations around the value $\theta = 0$. For $1/Ro > 0$, a linear, retrograde trend of $\theta$ is clear. The average retrograde rotation speed increases with increasing $1/Ro$. Since $\theta$ represents the orientation of the LSC with respect to a fixed point on the sample - i.e.\ as seen from the rotating sample frame - this linear trend and its increase with $1/Ro$ are unsurprising. They signify that the LSC - on average - rotates at a constant rate as well, but slower than the sample. As seen in the inset to Fig.~\ref{fig:Figure_2}, though, on short time-scales, this average trend is significantly distorted by fluctuations.

\begin{figure}[t!]
\centering
\includegraphics[width=1.0\textwidth]{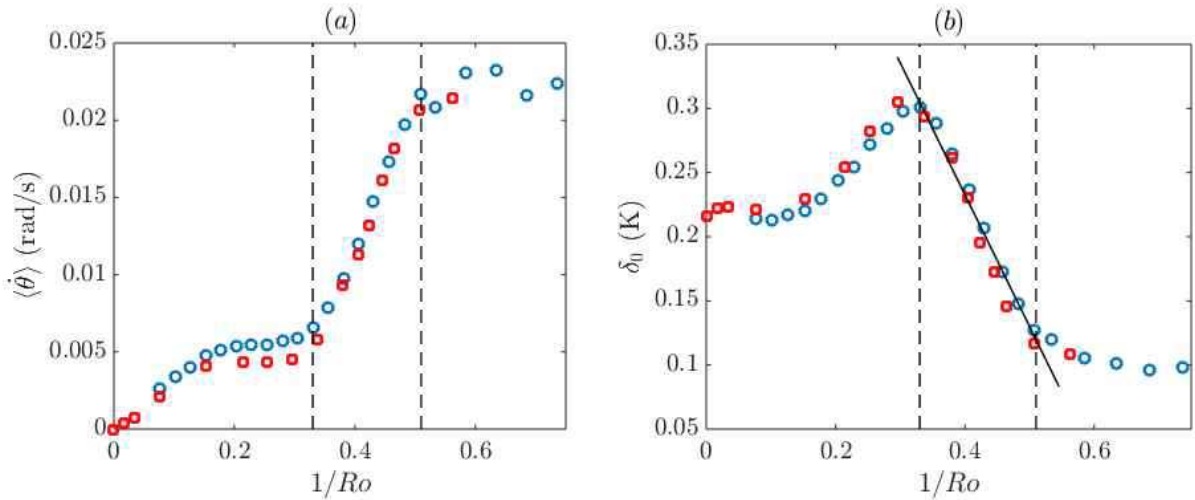}
\caption{Dynamical properties of the LSC when the sample rotates at constant rates. (a) The mean retrograde rotation velocity $\langle\dot{\theta}\rangle$ as a function of 1/Ro. Blue circles: experimental data from \cite{ZA10} with $Ra = 8.97\times10^9$; red squares: the present work with $Ra = 8.24\times10^9$. The range in which we perform modulated rotation experiment is indicated by the vertical dashed lines. (b) The mean LSC amplitude $\delta_0$ as a function of $1/Ro$. Solid line: linear fit to the squares from which we determine $\chi(\Omega) = -5.1/Ro + 3.1$ in equation~(\ref{eq:model}).}
\label{fig:Figure_3}
\end{figure}

We perform a linear fit to each of these curves to determine the mean retrograde rotation speed, denoted $\langle\dot{\theta}\rangle$, as a function of $1/Ro$. The result is given in Fig.~\ref{fig:Figure_3}(a). Beyond an initial increase with $1/Ro$, the curve levels off for $0.15 \lesssim 1/Ro \lesssim 0.30$, before increasing sharply for higher $1/Ro$. The same trend has been reported in~\cite{ZA10} (also included in Fig.~\ref{fig:Figure_3}). An explanation for the qualitative shape of this trend is currently unknown.

We also plot the time-averaged amplitude $\delta_0$ of the LSC as a function of $1/Ro$ in Fig.~\ref{fig:Figure_3}(b). We see that the average amplitude first increases with the inverse Rossby number, but reaches a peak and then drops sharply around $1/Ro \approx 0.3$. Apart from the variation in the temporal mean $\delta_0$, the time series of $\delta(t)$ do not exhibit significant differences for different values of $1/Ro$.

\subsection{Results for modulated rotation}

We chose $\Omega_0 = 0.104$~rad/s and $\beta = 0.212$, so the Rossby number $Ro = \sqrt{\alpha g \Delta T}/L$ varied periodically in the range $0.31 \leq 1/Ro \leq 0.5$ in the presence of modulation. As depicted in Fig.~\ref{fig:Figure_3}, in this parameter range (between the dashed vertical lines), the LSC retrograde rotation rate $\langle \dot{\theta} \rangle$ and its average thermal amplitude $\delta_0$ varied nearly linearly and most rapidly with $\Omega$, so we expected the strongest responses of these parameters to modulated values of~$\Omega$. The normalized modulation rate $\omega/\Omega_0$ ranged from~0 to~1.0.

The LSC flow velocity in its circulating plane, $U \approx 1.5$~cm/s (see also section~IV), was determined by approximating the turnover time of the LSC through the auto-correlation functions of the sidewall temperatures~\cite{ZA10}. Thus the Strouhal number $Sr = L\dot{\Omega}/(4\Omega U)$, which measured the ratio of the Euler force (the pseudo-force appearing in a frame of reference rotating at a time-dependent rate) and the Coriolis force, did not exceed 0.08~\cite{ZSL15}.

\begin{figure}[t!]
\centering
\includegraphics[width=0.55\textwidth]{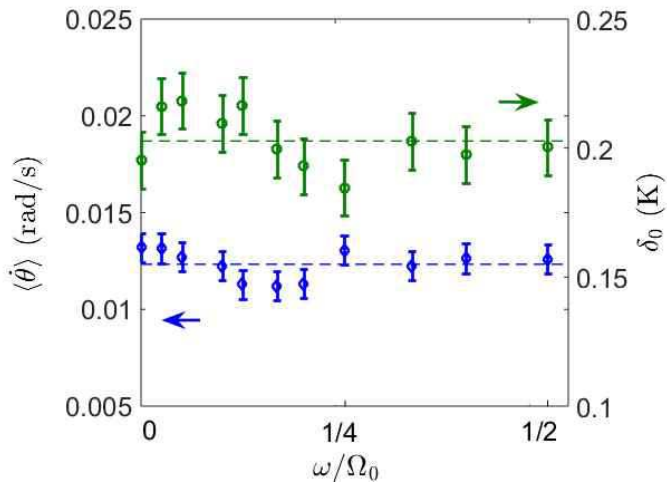}
\caption{Dynamical properties of the LSC when the sample rotation is modulated sinusoidally. (Left abscissa) The mean retrograde rotation velocity $\langle\dot{\theta}\rangle$ versus $\omega/\Omega_0$; (right abscissa) The mean LSC thermal amplitude $\delta_0$ versus $\omega/\Omega_0$. The dotted lines indicate the means across the range of $\omega/\Omega_0$; the error bars show the series' standard deviation with respect to this mean.}
\label{fig:Figure_4}
\end{figure}

In our experiments with modulated rotation, the orientation $\theta(t)$ of the LSC, as obtained from the cosine fitting procedure, is seen to exhibit a linear retrograde movement for all values of $\omega/\Omega_0$, just as in the constant-rotation experiments. In Fig.~\ref{fig:Figure_4}, we plot the linear retrograde rotation speed $\langle \dot{\theta} \rangle$ and the average thermal amplitude $\delta_0$ against $\omega/\Omega_0$. We have also included the experiment from the constant-rotation series ($\omega/\Omega_0 = 0$) that has the same mean $1/Ro = 0.42$. It is clear that neither $\langle \dot{\theta} \rangle$ nor $\delta_0$ is significantly affected by the modulation of the rotation rate.

\subsubsection{Modulation of azimuthal LSC velocity}

\begin{figure}[t!]
\centering
\includegraphics[width=0.9\textwidth]{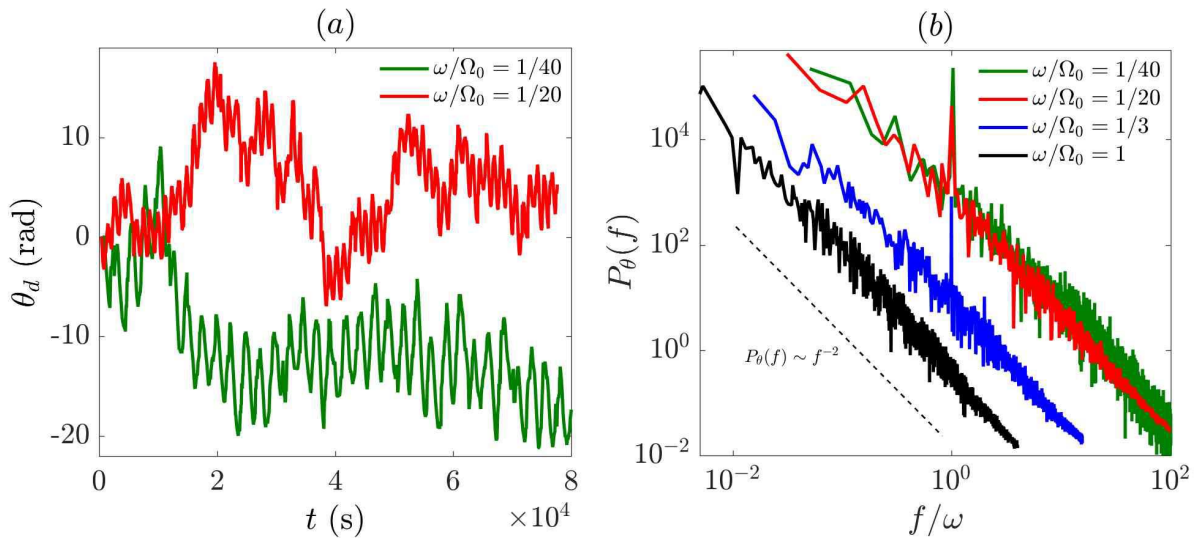}
\caption{(a) Two example series $\theta_{d}$ for different $\omega/\Omega_0$. The periodic behaviour in a noisy background is clear. (b) Power spectra of $\theta_d(t)$ for four different $\omega/\Omega_0$. There is a clear peak at the modulation frequency $\omega$ as long as $\omega < \omega_c$.}
\label{fig:Figure_5}
\end{figure}

It has been reported in \cite{XZX06, BA06a}, in the context of constant-rotation RB convection, that the fluctuations of $\theta$ around the linear retrograde trends have a diffusional character; i.e.\ the power spectrum of any detrended time series $\theta_{d} = -(\theta + \langle\dot{\theta}\rangle t)$ falls off with the frequency as a power law with exponent -2.
To establish the character of fluctuations of $\theta$ in the modulated-rotation case, we again calculate the detrended time series. In Fig.~\ref{fig:Figure_5}(a), we plot two example series $\theta_{d}$ for different $\omega/\Omega_0$. It is obvious in these plots that a periodic modulation in the LSC orientation can be seen once the linear retrograde trend is removed. There is thus a clear periodicity present in a noisy background. 

These example series correspond to very low modulation frequencies during which the response is extremely clear. We plot their power spectra $P_{\theta}$ in Fig.~\ref{fig:Figure_5}(b) against the corresponding normalized frequency $f/\omega$, along with two example power spectra for higher $\omega/\Omega_0$. The general fall-off slope of these curves is indeed consistent with $P_{\theta}(f) \sim f^{-2}$, as for constant rotation. It can be seen that the curves for $\omega/\Omega_0 = 1/40, 1/20$ exhibit a very clear peak at $f = \omega$, indicating a distinct presence of an oscillatory response in $\theta_d$; however, the peak becomes much weaker for $\omega/\Omega_0 = 1/3$ and disappears at $\omega/\Omega_0 = 1$. We find that this corresponds roughly to the critical value $\omega_c$, when the oscillatory response stops being distinguishable in the noisy time series of $\theta_d$. This could be explained by the fact that the modulation period $2\pi/\omega$ becomes smaller than the LSC turnover time $\mathcal{T} \approx \pi L/U \approx 50$~s for $\omega \gtrsim \Omega_0$~\cite{ZSL15}. 

\begin{figure}[t!]
\centering
\includegraphics[width=0.7\textwidth]{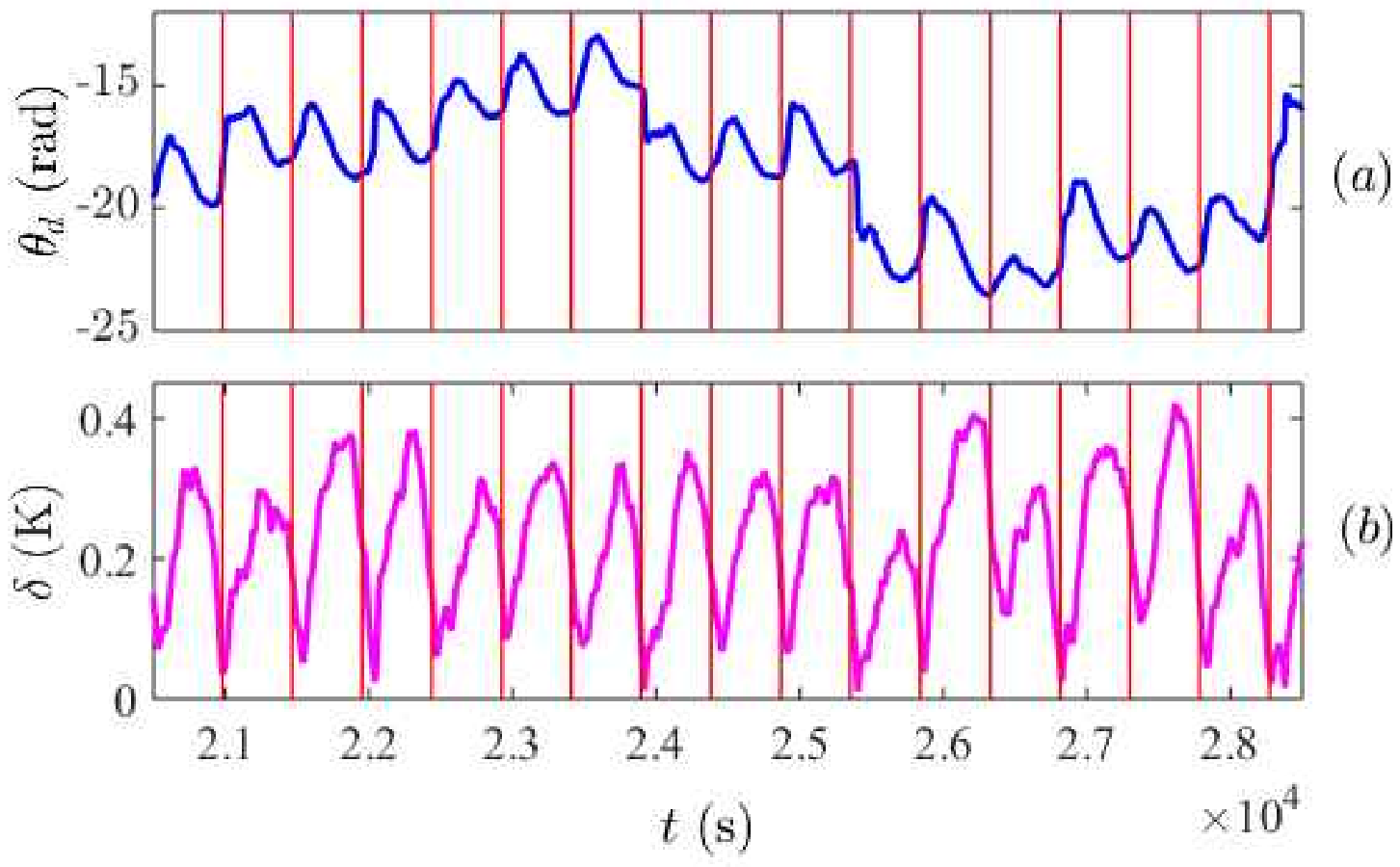}
\caption{Part of a time series for $\omega/\Omega_0 = 1/8$, showing the synchronization of the measured quantities $\theta_d(t)$ (a) and $\delta(t)$ (b). The vertical lines indicate the timings of maxima in the forcing $\Omega(t)$. Phase shifts between $\theta_d(t)$, $\delta(t)$ and $\Omega(t)$ thus become apparent.}
\label{fig:Figure_6}

\includegraphics[width=\textwidth]{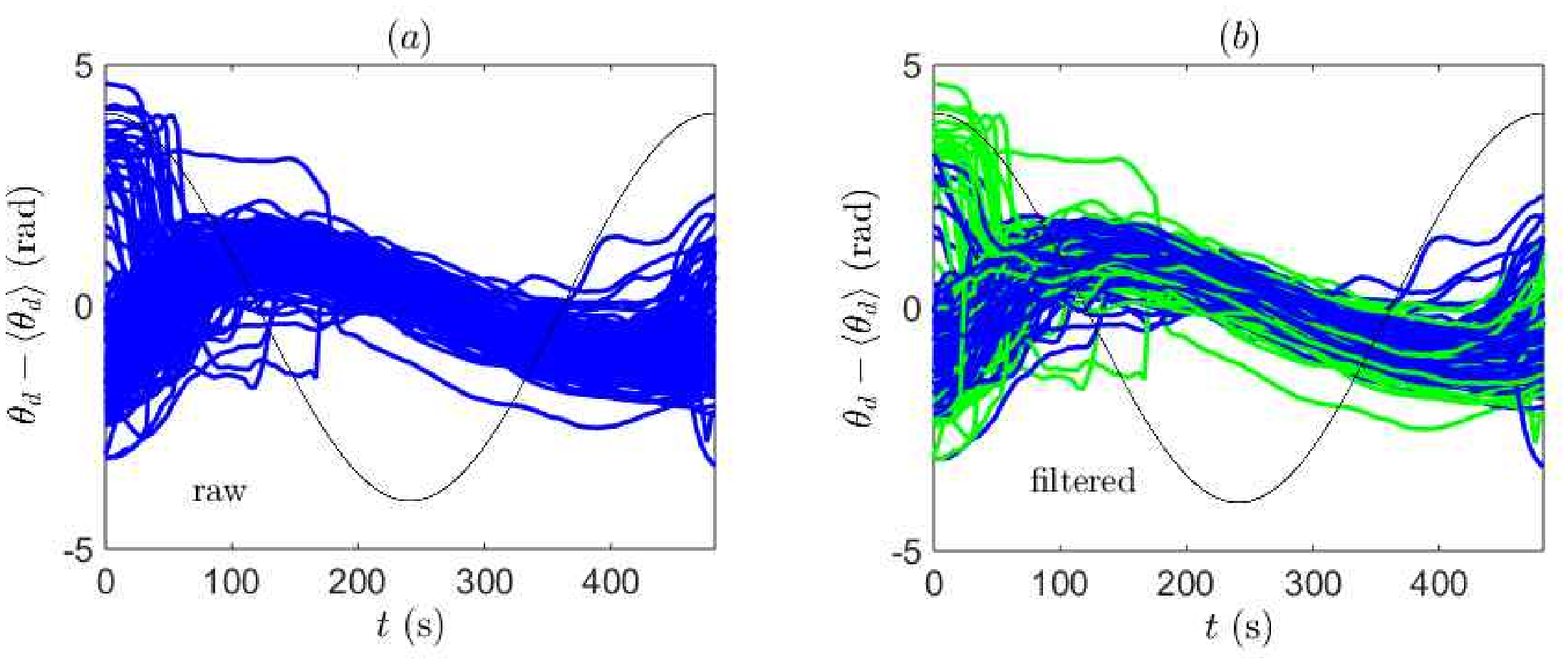}
\caption{(a) The ensemble of $\theta_d(t)$ for the same experiment as in Fig.~\ref{fig:Figure_6}. (b) Same as (a), but with the responses filtered out through the criteria mentioned in the text plotted in green, indicating how anomalous responses can be discarded from the ensemble. The smooth black curves represent $\Omega(t)$ in arbitrary units, showing clearly a phase difference between $\theta_d$ and $\Omega(t)$.}
\label{fig:Figure_7}

\end{figure}

In order to illustrate how the oscillations in LSC orientation are timed in comparison to the modulation of the RB cell, it is instructive to plot $\theta_d(t)$ and $\Omega(t)$ together. An example is given in Fig.~\ref{fig:Figure_6}(a), corresponding to an experiment with $\omega/\Omega = 1/8$. The vertical lines in this plot denote maxima in $\Omega(t)$. It is clear from this Figure that there is a well-defined phase shift of $\theta_d(t)$ with respect to $\Omega(t)$.
Even more instructive is to construct an ``ensemble oscillation'' of $\theta_d(t)$ and $\delta(t)$. This can be done by dividing the data into sections corresponding to one modulation period $T = 2\pi/\omega$ each, setting the mean of $\theta_d(t)$ in each of those sections to zero, and overlapping all the resulting curves for $\theta_d(t)$. 

\begin{figure}[t!]
\centering
\includegraphics[width=0.75\textwidth]{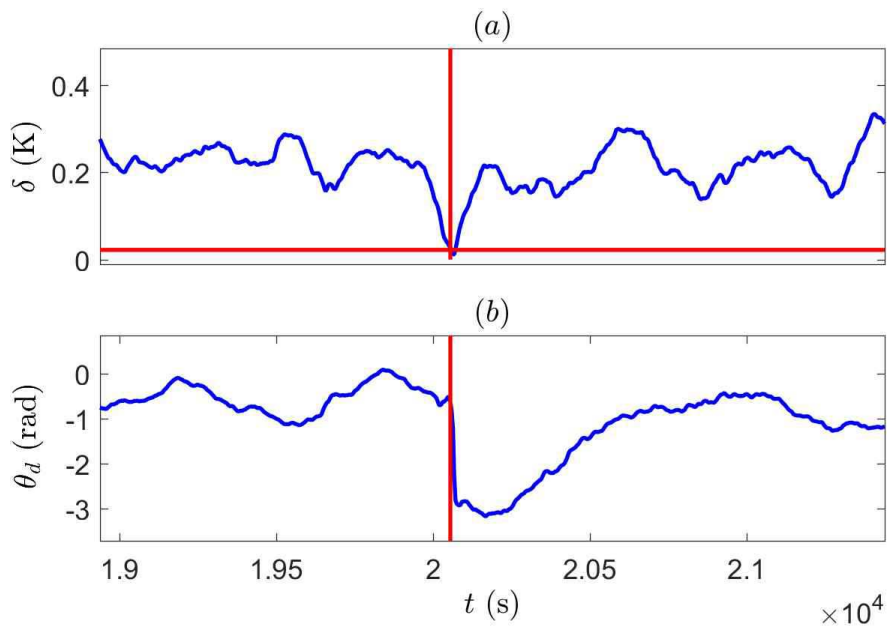}
\caption{Sudden drops in the thermal amplitude $\delta(t)$ of the LSC (a) are seen to be strongly correlated to sudden changes in $\theta_d(t)$ (b). This example data is for an experiment with $1/Ro = 0$, but such events occur also at finite rotation and modulation rates (see section~V and Fig.~\ref{fig:Figure_12}). The horizontal line indicates the criterion $\delta \leq 0.10\delta_0$; the vertical red lines indicate the moment where this criterion is first met, showing how it coincides with the sudden change in orientation.}
\label{fig:Figure_8}
\end{figure}

An example result, corresponding to the same experiment with $\omega/\Omega = 1/8$, is given in Fig.~\ref{fig:Figure_7}(a), which displays a well-defined ensemble oscillation. However, there are clear deviations from the ensemble oscillation as well, as can be clearly seen in the Figure. These are found to correspond to sudden changes in orientation of the LSC, and are generally correlated to very low values of $\delta(t)$. Such events happen when the LSC amplitude dips to near zero, stopping the overall circulation for a moment before it regenerates at a new orientation. We therefore classify these ``events'' as cessations, during which the LSC almost or completely vanishes \cite{BA06b}. An example (from an experiment without rotation, $1/Ro = 0$) of part of a time series of $\theta_d$ and $\delta$ containing a cessation event is shown in close-up Fig.~\ref{fig:Figure_8}, where the described characteristics can be clearly discerned. 

In order to correctly calculate the phase and amplitude responses of $\theta_d(t)$, these cessations need to be discarded from the ensemble, as $\theta_d$ does not display a clean oscillatory signal at these times. This was done as follows.
Since the cessations are strongly correlated to low values of $\delta(t)$, we firstly discard all the periods in which $\delta$ drops below $\delta_c \equiv 0.10\delta_0$ at least once. This criterion is based on the fact that the uncertainty in determining $\delta$ has a comparable magnitude to $\delta_c$; thus, such low values of $\delta$ are likely to represent the near-absence of an LSC.
Secondly, knowing that the criterion $\delta_c \equiv 0.10\delta_0$ does a good, but not a perfect job in filtering out event-affected periods, we also discard the other periods in which $\theta_d$ is so strongly affected by an event that its net rate of change $|\Delta\theta_d|/T$ from the start to the end of one modulation period ($T = 2\pi/\omega$) is larger than 0.01~rad/s. This criterion ensures that strongly deviating responses are filtered out, but at the same time that we do not discard responses in which the periodic behaviour could ``recover'' from an anomaly within one period $T$, which is mainly relevant for very slow modulations (where the periodic behaviour has enough time to recover from short-timescale reorientations for its phase and amplitude response to still be clearly measurable).

An example result is plotted in Fig.~\ref{fig:Figure_7}(b), where the periods discarded from Fig.~\ref{fig:Figure_7}(a) by the above criteria are plotted in green. It is seen that these criteria do a good job at ``cleaning up'' the data; nevertheless, they are not perfect. Thus, some unwanted signals due to cessations and possibly other events invariably do remain in the ensemble; however, their frequency of occurrence is extremely low, and therefore they no longer affect our data analysis.

From a physical point of view, we are more interested in the response of the azimuthal velocity $\dot{\theta}$ than the orientation itself. We thus set out to calculate $\dot{\theta}(t) = \partial \theta_d / \partial t $ from the raw data $\theta_d(t)$. For this, we smooth each data set $\theta_d(t)$ using a fourth-order Savitzky-Golay (SG) filter with a window length spanning one modulation period. In SG filtering of order $n$, a polynomial of order $n$ is fit to all points within a window; the value of this polynomial at the midpoint of this (odd-sized) window is taken to be the ``smoothed'' value at that point, and the value of the derivative of this polynomial at the midpoint is taken to be the derivative at that point. The window is then shifted by one point; the fitting is re-done, and the values at the next point are calculated. SG filtering can, of course, only be used to approximate derivatives up to the order of the filter itself.

\begin{figure}[t!]
\centering
\includegraphics[width=0.9\textwidth]{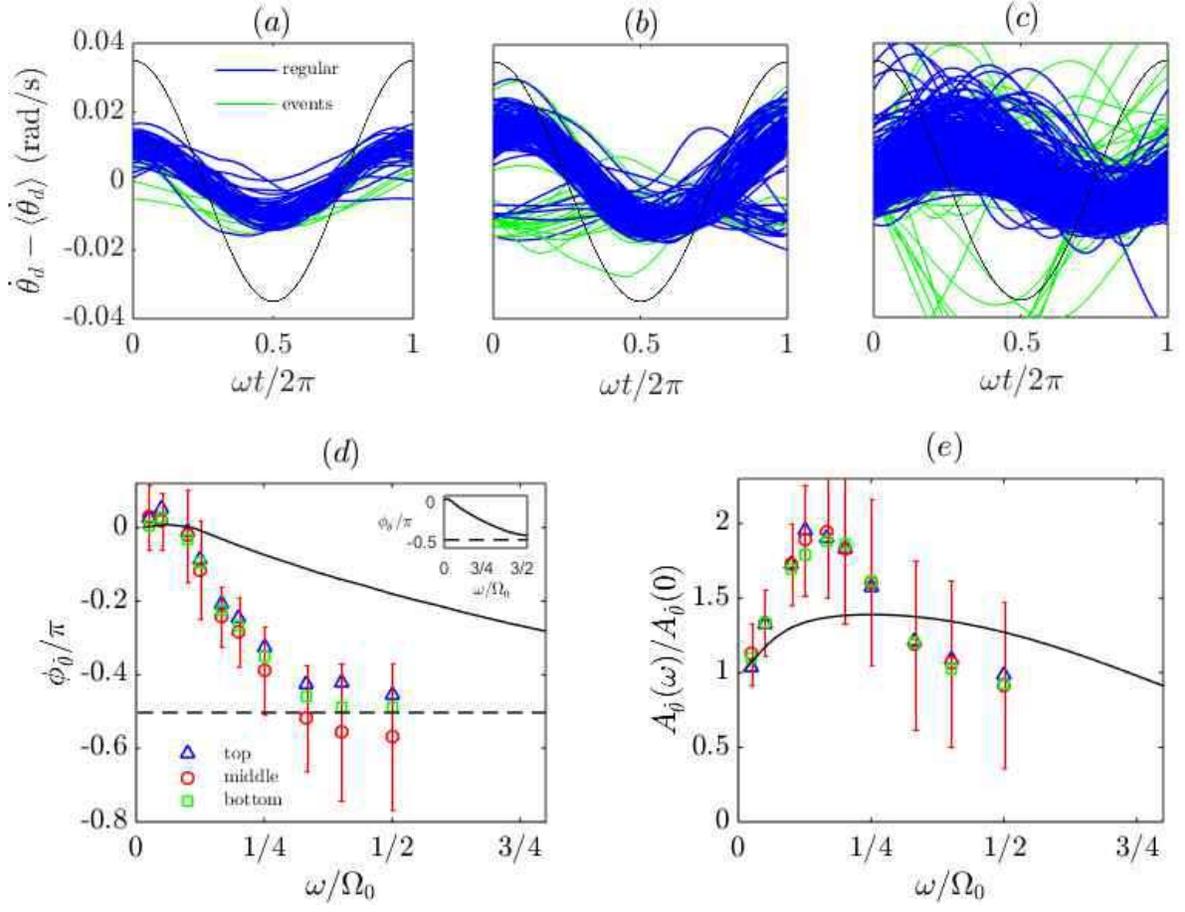}

\caption{(a)-(c): Ensembles of $\dot{\theta}_d$ for $\omega/\Omega_0 = 1/20, 1/8, 1/3$, respectively. The filtered-out responses are plotted in green, those kept are plotted in blue. The smooth black line is $\Omega(t)$ (in arbitrary units) to show the phase shift between $\dot{\theta}(t)$ and $\Omega(t)$. (d) Experimental results on the phase shift $\phi_{\dot{\theta}}$ as a function of $\omega/\Omega_0$, calculated from data from all three thermistor rows, and the corresponding numerical result (solid line) from equation~(\ref{eq:model}). The dotted line indicates a phase shift of $-\pi/2$, to which both experimental and model results converge for high $\omega/\Omega_0$. (Inset) The numerical result from equation~(\ref{eq:model}) with extended x-axis, to show its convergence to the same value $-\pi/2$ as experimentally observed. (e) Experimental results on the amplitude response $A_{\dot{\theta}}$ normalized by its value at zero modulation, $A_{\dot{\theta}}(0) = 0.010$~rad/s, as a function of $\omega/\Omega_0$, and the corresponding numerical result (solid line) from equation~(\ref{eq:model}).}
\label{fig:Figure_9}
\end{figure}

Using this method, we are able to discard the effects of noisy fluctuations and reliably estimate the oscillatory component of the azimuthal velocity.
In Fig.~\ref{fig:Figure_9}(a)-(c), we plot three examples of ensembles of $\dot{\theta}$ obtained in this way. From these ensembles, we can now directly calculate the phase shift $\phi_{\dot{\theta}}$ using a cross-correlation approach and taking the thermal diffusion time from fluid to thermistor into account (as explained in detail in Appendix B), and the amplitude response $A_{\dot{\theta}}$. The results (mean values with the error bars indicating standard deviations) are given in Fig.~\ref{fig:Figure_9}(d)-(e).

It is clear that $\phi_{\dot{\theta}}$ tends to $\approx-\pi/2$ as $\omega/\Omega_0$ increases. Furthermore, there is an initial increase of $A_{\dot{\theta}}$ with $\omega/\Omega_0$ followed by a decrease; the latter is to be expected in view of the fact that the oscillatory signal in $\theta_d$ gets lost for $\omega > \omega_c$. The maximum in $A_{\dot{\theta}}$ appears to represent a resonance. As explained later (see Section IV), a simple dynamical model coupling LSC orientation speed $\dot{\theta}$ to LSC strength $\delta$ can explain this as a resonant interaction between the LSC flow speed (which depends on its strength $\delta$) and the rotation speed of the sample, resulting in a Coriolis force with maximum amplitude at a finite $\omega$.

In Fig.~\ref{fig:Figure_9}(e), we have normalized $A_{\dot{\theta}}$ by the equivalent ``amplitude" of $\langle \dot{\theta} \rangle$ spanned in the relevant librational range. From Fig.~\ref{fig:Figure_3}(a), we estimate $\langle \dot{\theta} \rangle$ to vary by approximately 0.0149~rad/s between $1/Ro = 0.33$ and $1/Ro = 0.51$. Thus, if $\dot{\theta}_d$ only followed the average trend with $1/Ro$ without any lag, it would have an amplitude of roughly $A_{\dot{\theta},0} = 0.0149/2$~rad/s. Henceforth, we call this the ``adiabatic'' amplitude. (Correspondingly, there is also an adiabatic amplitude $A_{\delta,0}$ for $\delta(t)$.)
The limit of the quantity $A_{\dot{\theta}}/A_{\dot{\theta},0}$ for $\omega \rightarrow 0$ limit is indeed unity, as would be expected; the peak value of $A_{\dot{\theta}}/A_{\dot{\theta},0}$ is roughly twice as large.

\subsubsection{Modulation of LSC strength}

\begin{figure}[t!]
\centering
\includegraphics[width=0.7\textwidth]{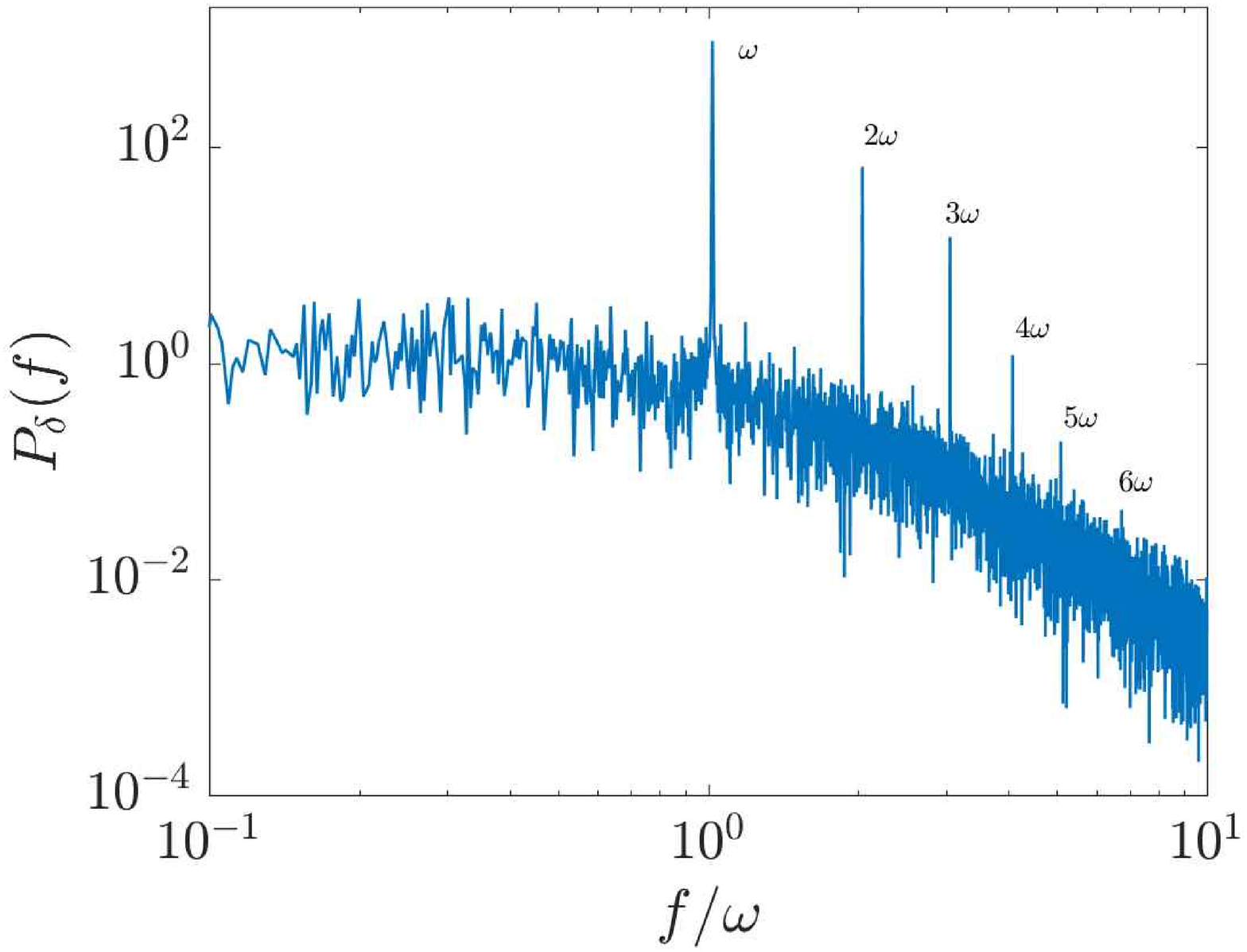}
\caption{An example power spectrum $P_{\delta}$ for $\omega/\Omega_0 = 1/10$.}
\label{fig:Figure_10}
\end{figure}

As already seen in Fig.~\ref{fig:Figure_4}, we find that the mean strength of the LSC $\delta_0$ is independent of $\omega/\Omega_0$. However, similar to $\theta_d(t)$, the amplitude $\delta(t)$ also contains a clear oscillation at the modulation frequency. An example of a synchronization plot of $\delta(t)$ with $\Omega(t)$ is given in Fig.~\ref{fig:Figure_6} (right abscissa). In Fig.~\ref{fig:Figure_10}, we show an example power spectrum $P_{\delta}$ from an experiment with $\omega/\Omega_0 = 1/10$. We see that $\delta(t)$ not only contains a dominant oscillation at frequency $f = \omega$, just like $P_{\theta}$. but also higher harmonics that are discernible (in this case) up to $f = 6\omega$, as indicated. As is the case for $\theta_d(t)$, the oscillatory signal for $\delta(t)$ gets weak at very high modulation rates, and disappears around~$\omega/\Omega_0 \approx 1$.

\begin{figure}[t!]
\centering
\includegraphics[width=0.9\textwidth]{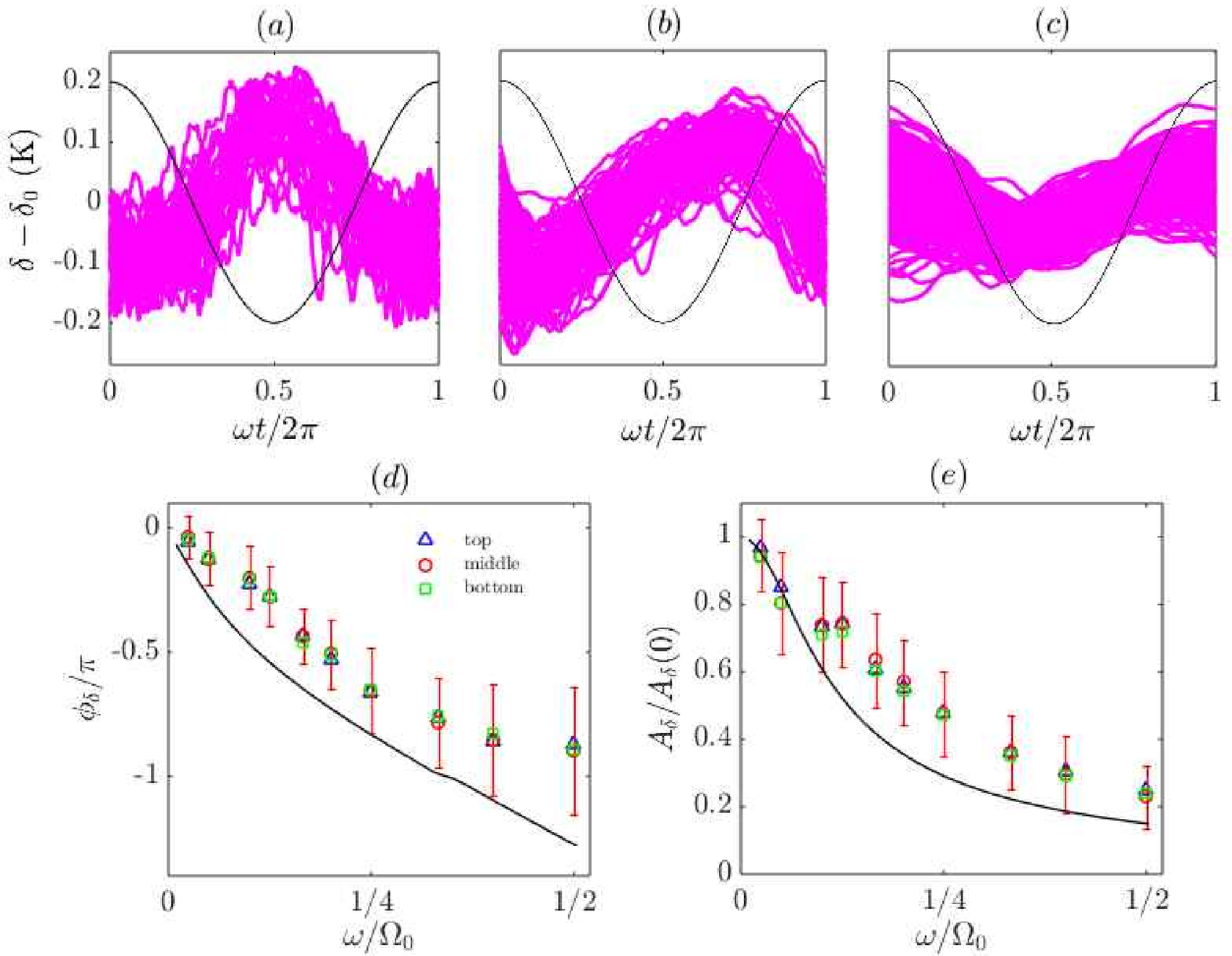}
\caption{(a)-(c): Ensembles of $\delta$ for $\omega/\Omega_0 = 1/40, 1/8, 1/3$, respectively. The smooth black lines show the shape of $\Omega(t)$ to define the phase relative to which $\phi_{\delta}$ is calculated. (d)-(e) The phase shift $\phi_{\delta}$ and the amplitude response $A_{\delta}$, respectively, as a function of $\omega/\Omega_0$, calculated from data from all three thermistor rows. The black line represents the numerical results from the model given by equation~(\ref{eq:model}).}
\label{fig:Figure_11}
\end{figure}

As was the case for $\theta_d(t)$, we can construct ensembles of $\delta(t)$ in exactly the same way. Three examples are given in Fig.~\ref{fig:Figure_11}(a)-(c). We note here that in Fig.~\ref{fig:Figure_11}(a), corresponding to $\omega/\Omega_0 = 1/40$, the slowest modulation rate investigated,~$\delta(t)$ looks to be in antiphase with $\Omega(t)$. This corresponds to the adiabatic response of $\delta$ to changes in $1/Ro$, since the dependence of $\delta$ on $1/Ro$ in the range $0.33 < 1/Ro < 0.51$ is approximately a linearly decreasing trend, cf.\ Fig.~\ref{fig:Figure_3}. We thus define the phase shift $\phi_{\delta}$ to be zero when $\delta(t)$ is in perfect antiphase to $\Omega(t)$.

Calculating the phase shift and the amplitude response from the ensembles of $\delta$ results in a mean and standard deviation for each $\omega/\Omega_0$; these two quantities are plotted in Fig.~\ref{fig:Figure_11}(d)-(e). We observe that the phase lag $\phi_{\delta}$ increases faster with $\omega/\Omega_0$ than does $\phi_{\dot{\theta}}$, with no apparent asymptotic limit for high $\omega/\Omega_0$; furthermore, the amplitude $A_{\delta}$ shows no maximum like $A_{\dot{\theta}}$, and decreases with $\omega/\Omega_0$ in a linear fashion. In the limit $\omega/\Omega_0 \rightarrow 0$, $A_{\delta}$ is seen to approach the adiabatic amplitude $A_{\delta,0}$, as should be expected.

\section{Modeling of the deterministic LSC dynamics}

In the following, we present an extended model of the LSC velocity and amplitude, based on earlier approaches by Brown \& Ahlers \cite{BA06a, BA07}, to explain the observed phase and amplitude responses. The basis of this model is formed by two Langevin-type equations for volume-averages of $\dot{\theta}$ and $\delta$. We first shortly explain the approach of Brown \& Ahlers to obtain these equations in the context of constant-rotation RB convection, before extending the model to include the effects of modulated rotations. Results from this model have been previously described in \cite{ZSL15} in less detail and are presented with more comprehensive explanations here.

The Langevin equation for $\delta$ is obtained starting from the Navier-Stokes (NS) equation in the polar direction, keeping buoyancy and drag terms: $\dot{u}_{\phi} = g\alpha(T-T_0) + \nu\nabla^2 u_{\phi}$. Performing a suitable volume averaging, assuming that the temperature and velocity profiles are linear in the radial coordinate, and assuming that the polar velocity is instantaneously proportional to the thermal amplitude (for details, see \cite{BA06a}), the equation for $\delta$ becomes

\begin{equation}
	\dot{\delta} = \frac{\delta}{\tau_{\delta}} - \frac{\delta^{3/2}}{\tau_{\delta}\delta_0^{1/2}},
	\label{eq:delta}
\end{equation}
with $\tau_{\delta} = L^2/(18\nu Re_{m}^{1/2})$ and $\delta_0 = 18\pi\Delta T Pr Ra^{-1} Re_{m}^{3/2}$. Here, $Re_{m}$ is the time-averaged Reynolds number $UL/\nu$. For our experimental parameters, $\tau_{\delta} \approx 62$~s and $\delta_0 \approx 0.22$~K (for the latter, cf.\ also the experimental results in Fig.~\ref{fig:Figure_4}).

The Langevin equation for $\dot{\theta}$ is obtained starting from the NS equation in the azimuthal direction, keeping rotational pseudo-forces and viscous drag: $\dot{u}_{\theta} = -2(\Omega + \dot{\theta})\times u_{\phi} - \ddot{\theta} \times r + \nu \nabla^2 u_{\theta}$. We assume that the Euler acceleration $\sim \dot{\Omega} \times r$ is much smaller than the Coriolis acceleration, and can be neglected in this approach (see section~III-B).
Again, assuming that the velocity profiles are linear in the radial coordinate, performing a suitable volume averaging (for details, see again \cite{BA06a}), employing the same proportionality between polar velocity and thermal LSC amplitude as in the derivation of equation~(\ref{eq:delta}), and lastly defining the direction of $\dot{\theta}$ to be prograde (to ensure comparability with the experimental results, where $\theta_d = -(\theta + \langle \dot{\theta}\rangle t)$ is prograde), the equation for $\dot{\theta}$ becomes

\begin{equation}
	\ddot{\theta} = -\left(\frac{\delta}{\tau_{\dot{\theta}}\delta_0} + \frac{\delta^{1/2}}{2\tau_{\delta}\delta_0^{1/2}} \right)\dot{\theta} + \frac{\delta}{\tau_{\dot{\theta}}\delta_0}\Omega,
	\label{eq:thetadot}
\end{equation}
where $\tau_{\dot{\theta}} = 4L^2/(3\nu Re_{m})$; for our experimental parameters, $\tau_{\dot{\theta}} \approx 19$~s.

It is clear that a modulated rotation rate $\Omega(t)$ will result in a modulated response of $\dot{\theta}$ in this model, due to the modulation of the Coriolis term $\sim \Omega$. However, the equation for $\delta$ does not contain any terms that respond to a temporal change of $\Omega$. This has to be amended by taking into account the $\Omega$-dependence of the momentum BL thickness~$\lambda$ (see i.e.\ \cite{SCL10}), which modifies the viscous drag terms in both equations~(\ref{eq:delta}) and~(\ref{eq:thetadot}). Physically, it means that the thickness of the viscous boundary layers will periodically change along with the rotation rate of the RB cell, resulting in a periodically modulated drag force.

Based on arguments by Assaf et al.\ \cite{AAG12}, the $\Omega$-dependence of the momentum BL thickness can be quantified as $\chi(\Omega) \equiv \lambda^2(\Omega)/\lambda^2(0) \approx \delta(\Omega)/\delta(0)$, the latter of which can be directly obtained from the experimental result shown in Fig.~(\ref{fig:Figure_3}). The viscous drag terms in both equations, which depend on $\lambda$ as $\sim 1/\lambda$ (cf.\ \cite{BA06a}), then have to be multiplied by $\chi(\Omega)^{-1/2}$.

Furthermore, we assume that it takes a finite time for the bulk circulation to respond to the modulation of the BL thickness, which should be of the order of the LSC turnover time $\mathcal{T} \approx \pi L/U \approx 50$~s. This effect is included in the model by using $\Omega^{*}(t) = \Omega(t - \mathcal{T})$, instead of $\Omega(t)$, to calculate the time-dependent drag terms.

The full system of equations thus becomes

\begin{equation}
		\begin{cases} 
			 \dot{\delta} = \dfrac{\delta}{\tau_{\delta}} - \dfrac{\delta^{3/2}}{\tau_{\delta} \delta_0^{1/2} \sqrt{\chi(\Omega^*)}}; \\
			\\
				\ddot{\theta} = \mbox{ }-\left(\dfrac{\delta}{\tau_{\dot{\theta}}\delta_0} + \dfrac{\delta^{1/2}}{2\tau_{\delta} \delta_0^{1/2} \sqrt{\chi(\Omega^*)}} \right)\dot{\theta} + \dfrac{\delta}{\tau_{\dot{\theta}}\delta_0}\Omega. \\
		\end{cases}
		\label{eq:model}
\end{equation}

The only free parameter in this model is the typical LSC flow speed $U$, which is contained in the time constants $\tau_{\delta}$ and $\tau_{\dot{\theta}}$ as well as in $\delta_0$. We use the value $U = 1.5$~cm/s (see section III-B) which is typical for the $Ra$, $Pr$ and $\Gamma$ values with which our experiments are concerned. We now compare the predictions by the model to the experimentally obtained results. All system parameters ($L$, $\nu$, etc.) in the model are thus taken equal to those used in our experiments. We solve the system~(\ref{eq:model}) using numerical integration with first-order time stepping.

Results for $\phi_{\dot{\theta}}$ and $A_{\dot{\theta}}$ are given in Fig.~\ref{fig:Figure_9}(d)-(e). Here, it can be seen that model and experiment are in qualitative agreement: the model reproduces both the asymptotic value of $-\pi/2$ for the phase shift of $\dot{\theta}$ at large $\omega/\Omega_0$, as well as the maximum at finite $\omega/\Omega_0$ for $A_{\dot{\theta}}$. The range of $\omega/\Omega_0$ in which these developments are projected to happen (top horizontal axis) is, however, larger than measured experimentally (bottom horizontal axis) in both cases. We assume that this is due to a relative underestimation of the strength of the azimuthal fluid acceleration of the LSC in equation~\ref{eq:model} (the term $\sim \ddot{\theta}$) in comparison to the inertial and viscous terms. 

The model also provides an explanation for the resonant peak in $A_{\dot{\theta}}$ observed experimentally. This peak is caused by an optimal coupling between $\delta(t)$ and $\Omega(t)$ in the Coriolis term $\sim {\delta}/(\tau_{\dot{\theta}}\delta_0)\Omega$. As both $\delta(t)$ and $\Omega(t)$ are oscillating functions with a phase shift $\phi_{\delta}(\omega) + \pi$ between them, and they are in perfect antiphase in the limit $\omega \rightarrow 0$, the amplitude of the Coriolis term reaches a maximum at a finite $\omega/\Omega_0$.

Results for $\phi_{\delta}$ and $A_{\delta}$ have been plotted in Fig.~\ref{fig:Figure_11}(d)-(e). Here, the model agrees both qualitatively and quantitatively very well with the experimental results, showing both the strong phase lag as well as the continuously decreasing amplitude of $\delta$ with increasing $\omega/\Omega_0$.

\section{Experimental results: Statistical dynamics of the LSC flow}

In this section, we provide in-depth results on the influence of modulated rotation rates on the \textit{statistical} behaviour of cessation events and the way in which these influence the overall statistics of $\dot{\theta}$ and $\delta$. The results discussed here pertain to the same parameter ranges as in the previous two sections and are obtained from the same experimental runs and/or repeats thereof.

\subsection{Cessation frequency}

\begin{figure}[t!]
\centering
\includegraphics[width=0.85\textwidth]{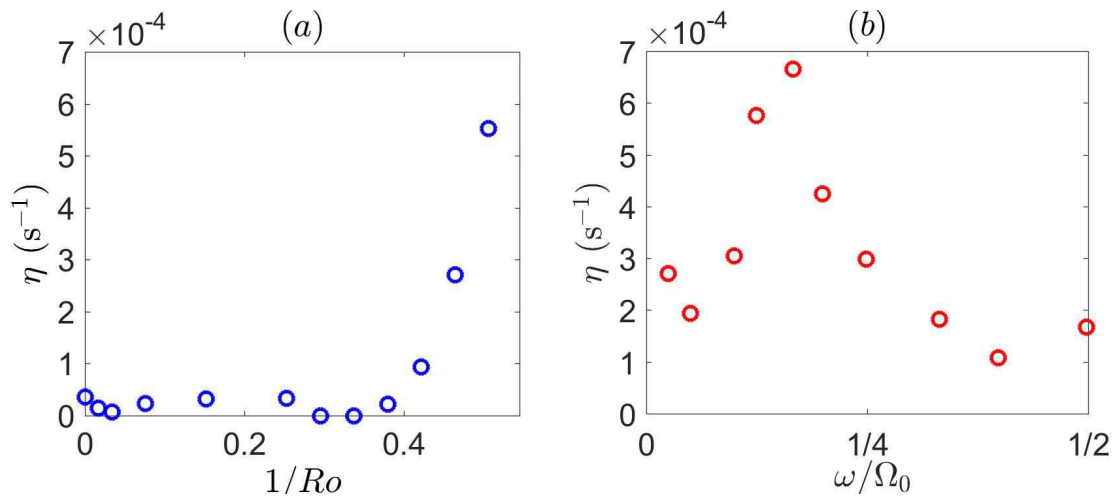}

\caption{Measured frequency of cessation events by the criterion $\delta < \delta_c \equiv 0.10\langle\delta\rangle$ as a function of (a) $1/Ro$ and (b)~$\omega/\Omega_0$.}
\label{fig:Figure_12}
\end{figure}

In section~III-B, we have mentioned the identification of cessations by the criterion $\delta < \delta_c \equiv 0.10\langle\delta\rangle$. In the context of constant-rotating RB convection, we find that the frequency of cessations $\eta$ increases rapidly beyond $1/Ro \approx 0.40$, as also reported before for comparable $Ra$ in~\cite{ZA10}. The dependence of $\eta$ on $1/Ro$ as measured in our current study is plotted in Fig.~\ref{fig:Figure_12}(a).
Interestingly, in our experiments with modulated rotation rates, we also find a nontrivial dependence of~$\eta$ on $\omega/\Omega_0$, plotted in Fig.~\ref{fig:Figure_12}(b). There appears to be a maximum in~$\eta$ around $\omega/\Omega_0 \approx  1/6$.

We have repeated two experiments from the modulated-rotation series, namely those with $\omega/\Omega_0 = 1/10, 1/6$, for a duration of approximately an entire week each. We note that these two values of $\omega/\Omega_0$ have been chosen on the basis of their proximity to the peak in cessation frequency (Fig.~\ref{fig:Figure_12}), enabling us to make statistical inferences about the cessation events themselves, and the dependence of those statistics on the phase of oscillation.

\begin{figure}[t!]
\centering
\includegraphics[width=0.9\textwidth]{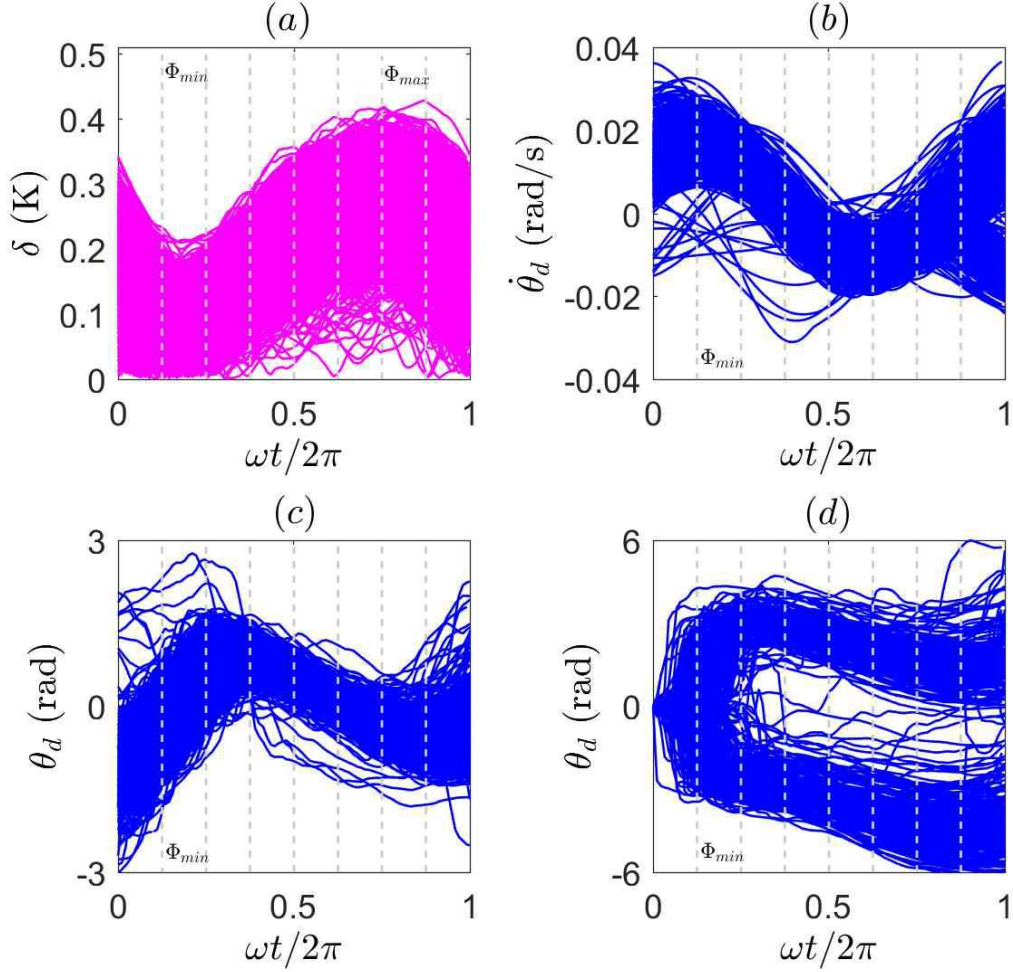}
\caption{The ensembles of (a) $\delta$, (b) $\dot{\theta}$ (clean response), (c) $\theta_d$ (clean response), and (d) Event-affected responses of $\theta_d$, shifted to have $\theta_d(0) = 0$ for the sake of clarity; corresponding to the $\omega/\Omega_0 = 1/6$ experiment. $\Phi_{min}$ denotes the phase in which $\delta(t)$, on average, reaches the minimum values of its periodic response, and in which events thus have the highest probability of occurring; similarly, $\Phi{max}$ denotes the phase where $\delta(t)$ reaches its maximum values.}
\label{fig:Figure_13}
\end{figure}

In Fig.~\ref{fig:Figure_13}(a)-(c), we show the ensembles of $\delta$ (all responses), $\dot{\theta}$ (without events), and $\theta_d$ (without events), respectively, for the $\omega/\Omega_0 = 1/6$ run, which is near the maximum in $\eta$. The vertical lines here indicate the division of one period $T = 2\pi/\omega$ into $n$ phases, denoted~$\Phi_n$ (in the Figure, $n = 8$). Since cessations are identified by near-zero values of $\delta$, it is easy to see how the modulation of $\delta$ tends to ``concentrate'' the cessations in a certain phase which we denote $\Phi_{min}$, where $\delta$ reaches the minimum values of its periodic response (indicated in Fig.~\ref{fig:Figure_13}).

\begin{figure}[t!]
\centering
\includegraphics[width=0.8\textwidth]{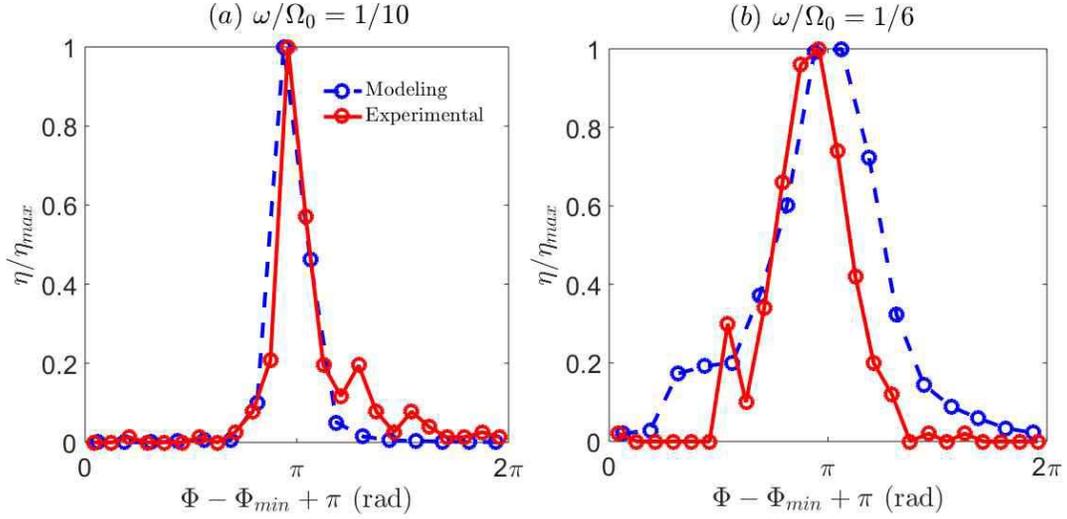}
\caption{(a) The normalized frequency of cessations for each individual phase $\Phi_n$ determined from data of the long experimental run with $\omega/\Omega_0 = 1/10$; experimental (solid) and model (dashed) results. (b) Same for $\omega/\Omega_0 = 1/6$. The curves show how equation~(\ref{eq:master_cess}) roughly reproduces the experimentally measured dependency.}
\label{fig:Figure_14}
\end{figure}

We recorded more than~300 event-affected responses in this experiment. This enabled us to construct a representative \textit{ensemble of events}. In Fig.~\ref{fig:Figure_13}(d), we give such an ensemble for $\theta_d(t)$. For the sake of clarity, we have shifted each of these curves to be zero at $t = 0$. It can be seen that these anomalous responses are manifested as distinct, rather abrupt changes in orientation of the LSC circulation in both directions, and that most of the anomalies in $\theta_d$ are concentrated inside the phase $\Phi_{min}$. 
This concentration of cessations can be illustrated by showing the frequency of cessations for each individual phase~$\Phi_n$. This quantity is plotted in Fig.~\ref{fig:Figure_14}(a)-(b) for the two different experiments, respectively, using $n = 24$ (series ``Experimental"). Here, the horizontal axis has been shifted by the phase corresponding to $\Phi_{min}$, to harmonize the plots for $\omega/\Omega_0 = 1/10$ and $\omega/\Omega_0 = 1/6$ ($\Phi_{min}$ changes with $\omega/\Omega_0$ because $\phi_{\delta}$ changes with $\omega$). It can be seen that the curves are roughly symmetrical and exhibit a very sharp peak among the 24~phases. The cessations thus have a very high probability of occurring in a very small phase window, and during the rest of each period $T$, the circulation is nearly always sustained. It can also be seen that this phase window is broader for the higher $\omega/\Omega_0$. In section~V, we provide a theoretical model for the shape of~$\eta(\Phi)$.

\subsection{Probability distributions of $\delta$ and $\dot{\theta}$}

\begin{figure}[t!]
\centering
\includegraphics[width=0.8\textwidth]{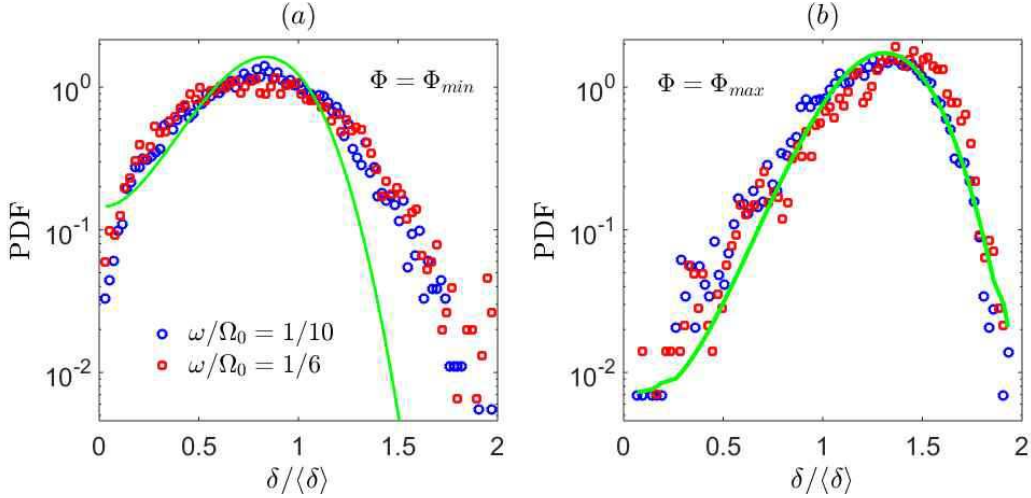}
\caption{The probability distribution function (PDF) of $\delta$ in (a) the phase $\Phi_{min}$ during which the ensemble mean of $\delta$ is minimal, and (b) the phase $\Phi_{max}$ during which the ensemble mean is maximal, for two different $\omega/\Omega_0$.}
\label{fig:Figure_15}
\end{figure}

Clearly, a number of statistical properties of LSC dynamics will depend on the phase $\Phi$. In Fig.~\ref{fig:Figure_15}, we plot the probability distribution function (PDF) of $\delta$ (normalized by its mean) in the phase $\Phi_{min}$ during which the ensemble mean is minimal (i.e.\ where the frequency of cessations is maximal), and in the phase $\Phi_{max}$ during which the ensemble mean of $\delta$ is maximal, for both $\omega/\Omega_0 = 1/10$ and $\omega/\Omega_0 = 1/6$. This Figure illustrates clearly the different skewness of $\delta$ in different phases - it is clear how for $\Phi = \Phi_{min}$, the points in the left tail of the PDF are bunched together closely (near $\delta = 0$, which provides an absolute constraint as $\delta$ cannot be negative), thus giving the PDF a very different shape as compared to $\Phi = \Phi_{max}$, where such low values are nearly never reached in the absence of cessations. It can also be seen that the normalized PDFs for the different $\omega/\Omega_0$ overlap to a large extent. 

The effect of the cessations on the reorientation of the LSC circulation plane can be illustrated by the PDFs of the absolute angular change $|\dot{\theta}|$. To calculate $\dot{\theta}$, now, we should not use the SG filtering approach detailed earlier, since it would smooth out the short-timescale effects of cessations; rather, we calculate $\dot{\theta}(t) = (\theta_d(t + \Delta t) - \theta_d(t))/\Delta t$, with $\Delta t = 4$~s, the temporal resolution of our data recordings. Cessations will then be manifested by anomalously high values of $|\dot{\theta}|$; thus, the PDFs of $|\dot{\theta}|$ for different $\Phi$ can provide us with more information on the effects of cessations on LSC statistics, and their dependence on the modulation phase.

\begin{figure}[t!]
\centering
\includegraphics[width=\textwidth]{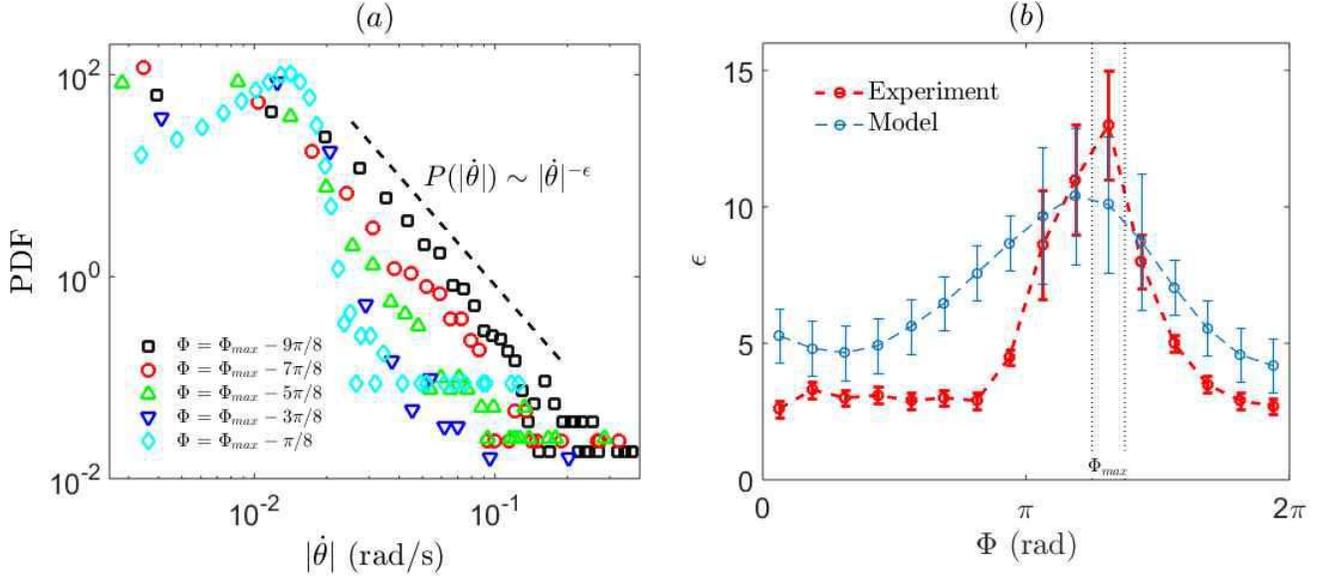}
\caption{(a) The PDFs of $|\dot{\theta}|$ for $\omega/\Omega_0 = 1/10$ in a number of different phases. (b) The fall-off of the PDFs at high $|\dot{\theta}|$ can be approximated by a power law, $P(|\dot{\theta}|) \sim |\dot{\theta}|^{-\epsilon}$; here, we plot $\epsilon(\Phi)$ from experimental results (left abscissa) and from the modeling approach of equation~(\ref{eq:pdottheta}) (right abscissa). Error bars indicate estimates of the uncertainty of the result of the exponential fitting by shifting the range in which the fits are performed.}
\label{fig:Figure_16}
\end{figure}

In Fig.~\ref{fig:Figure_16}a, we plot the PDF of $|\dot{\theta}|$ for $\omega/\Omega_0 = 1/10$ in a number of different phases. The fall-off of the PDFs at high $|\dot{\theta}|$, the regime where cessations become dominant, can be approximated by a power law, $P(|\dot{\theta}|) \sim |\dot{\theta}|^{-\epsilon}$.
It is evident that $\epsilon = \epsilon(\Phi)$. We have estimated $\epsilon$ by a suitable fit in the PDF tail for all curves; we plot $\epsilon(\Phi)$ in Fig.~\ref{fig:Figure_16}(b). It is clear that during $\Phi_{max}$, when cessations almost never occur, $\epsilon$ reaches extremely high values compared to other $\Phi$, in which cessations are more common.

\subsection{Interplay between cessations and LSC dynamics}

\begin{figure}
\centering
\includegraphics[width=0.8\textwidth]{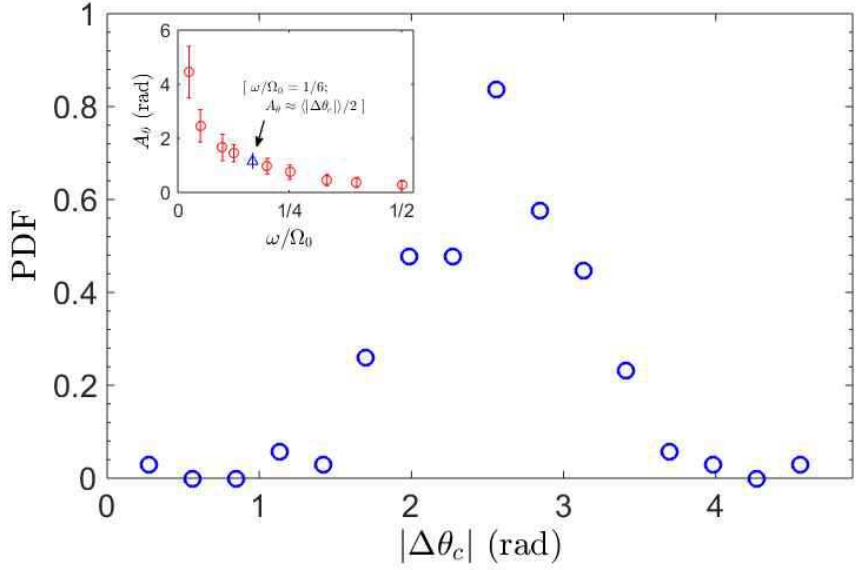}
\caption{The probability distribution of the change in LSC orientation $|\Delta\theta_c|$ following a cessation, for the long $\omega/\Omega_0 = 1/6$ experiment. Inset: the amplitudes $A_{\theta}$ as a function of $\omega/\Omega_0$; the value at which $\langle |\Delta\theta_c| \rangle \approx 2\langle A_{\theta} \rangle$ is indicated by a triangular data point. This value happens to correspond to $\omega/\Omega_0 = 1/6$.}
\label{fig:Figure_17}
\end{figure}

In Fig.~\ref{fig:Figure_17}, we plot the probability distribution of the magnitude of the angular change during cessations, which we denote $|\Delta\theta_c|$, for the $\omega/\Omega_0 = 1/6$ experiment. The results from this graph, however, apply not only to $\omega/\Omega_0 = 1/6$ but turn out to be accurate across a wide range of $\omega/\Omega_0$. The explanation for this is as follows. The mean angular change during cessations is given by $\langle | \Delta\theta_c |\rangle = \sqrt{\tau D_{\theta,c}}$. Here $\tau$ is the mean duration of cessation, and $D_{\theta,c}$ is the mean ``effective" diffusivity of $\theta_c$ during cessations. Now both $\tau$ and $D_{\theta,c}$ are independent of the modulation as long as the period of modulation is much longer than $\tau$. (One can estimate the order of these quantities, for instance, from Fig.~\ref{fig:Figure_8}, where $|\Delta\theta_d| \approx 2.4$~rad and $\tau \approx 12$~s, giving $D_{\theta,c} \sim 0.5$~rad$^2$/s.) Thus, Fig.~\ref{fig:Figure_17} can be seen as being representative for the modulated-rotation experiments in general.

The mean value of $|\Delta\theta_c|$ is seen to be approximately~2.5~rad. The amplitude $A_{\theta}$, meanwhile, is found to be $1.2 \pm 0.3$~rad for this run (cf.\ Fig.~\ref{fig:Figure_13}(c) and the inset in Fig.~\ref{fig:Figure_17}). Thus, $\langle |\Delta\theta_c| \rangle \approx 2\langle A_{\theta} \rangle$, i.e.\ on average, the magnitude of a sudden change in LSC orientation is roughly equal to twice the ``clean'' ensemble amplitude $A_{\theta}$, or equal to the mean peak-to-peak variation of $\theta$ within one period. 

As has been discussed, the sudden changes in LSC orientation are correlated to minima in~$\delta$. For this $\omega/\Omega_0$, the phase in which minima in $\delta$ are concentrated ($\Phi_{min}$) happens to also coincide with the phase where $\dot{\theta}_d$ is largest (cf.\ Fig.~\ref{fig:Figure_13}(b)). The implication is that, whenever an event occurs that manifests itself as an anomalous change in LSC orientation, this has a very high probability of happening in the same phase in which the ``clean'' signal $\theta_d$ would otherwise have exhibited its fastest change. 

Thus, near this value $\omega/\Omega_0 = 1/6$, the angular change due to cessations (Fig.~\ref{fig:Figure_13}(d)) tend to synchronize with the clean ensemble response (Fig.~\ref{fig:Figure_13}(c)), because they span roughly the same angular range, and match closely in phase with the clean response. These two factors provide the conditions for a kind of resonance: most sudden changes in orientation do not interrupt the oscillatory ensemble response of $\theta_d$, as they do for other values of $\omega/\Omega_0$, but instead are obscured within the time series by having roughly the same amplitude and phase. The result is a time series that exhibits almost the same oscillation pattern during each subsequent phase, unchanged - in fact, even enhanced - by the presence of cessations. 

This enhances a number of physical properties of the flow. For example, while the LSC orientation oscillates about its mean value much more regularly than for other $\omega/\Omega_0$ - where cessations interrupt the flow instead of strengthening its pattern - the thermal amplitude~$\delta$ oscillates about its mean value as usually. This implies that the LSC leaves its ``footprint" (presumably, manifested by the presence of traces of cold or hot fluid near the sides of the cylinder walls) behind in a much more spatially regular pattern than for other $\omega/\Omega_0$. The ``footprint" of the minimum phase (in which there is a much smaller temperature difference between fluid carried up/downwards on opposite sides of the sample than in other phases) is therefore reinforced in the same spatial location as seen from the rotating cylinder during each period of the modulation. 

Since cessations are defined by anomalously low values of $\delta$, the question now becomes: why do such anomalously low values occur so often at $\omega/\Omega_0 \approx 1/6$? 
This could be down to a resonant effect, resulting from the enhanced spatial regularity of the flow. Consider that traces of the previous minimum phase are still present somewhere in the sample during the next modulation period (perhaps as a small temperature anomaly in the thermal BL). If $\omega/\Omega_0 \approx 1/6$, then during the next minimum phase, when the LSC would be in the same location in the rotating frame as during the previous one, this could lead to a slightly higher chance of $\delta$ dropping low enough for a cessation to occur than if the LSC had been in any other position during the same phase. The same process would be repeated again during the subsequent minimum phase, until a cessation would indeed occur at some point. 

Now, as we have seen, the cessation, due to its close amplitude matching with the clean ensemble response, would effectively not interrupt the modulated flow pattern of the LSC. Thus, the process of a previous minimum phase reinforcing the next one would continue unabated afterwards. For other $\omega/\Omega_0$, an LSC regenerated after cessation would be in a different position as compared to where it would have been had the cessation not occurred; thus, this process of reinforcement would be interrupted after any cessation. At $\omega/\Omega_0 \approx 1/6$, there is no such constraint. 

This sets the stage for a resonant effect in which cessations, normally stochastic processes, become more likely to happen during each subsequent modulation cycle, resulting in anomalously high and regular occurrences of cessations at $\omega/\Omega_0 \approx 1/6$. We may notice that the peak in $\eta$ occurs roughly at the same $\omega/\Omega_0$ as the peak in $A_{\dot{\theta}}$. While we have removed the effect of cessations in the analysis of $A_{\dot{\theta}}$ by removing all periods containing cessations from the analysis, a resonance such as theorized above could still contribute by a small amount to the peak in $A_{\dot{\theta}}$. This could help explain why the model, despite predicting the occurrence of the peak in $A_{\dot{\theta}}$, underestimates its magnitude.

\section{Theory of the stochastic LSC behavior}

In this section, we provide theoretical explanations for the statistical phenomena observed in our experiments as detailed in section~V. It is seen that the LSC model in equation~(\ref{eq:model}) has provided reasonable predictions of the deterministic dynamics of the LSC flows subjected to modulated rotations. In order to describe the stochastic behavior of the LSC, i.e.\ the statistics of cessation events, the probability distributions of both $\dot{\theta}(t)$ and $\delta(t)$, and particularly their dependence on the modulation phases, we consider an extended theory with stochastic terms included that model the small-scale turbulent fluctuations in the
fluid background.

The stochastic system of equations is then given by
\begin{equation}
		\begin{cases} 
			 \dot{\delta} = \dfrac{\delta}{\tau_{\delta}} - \dfrac{\delta^{3/2}}{\tau_{\delta} \delta_0^{1/2} \sqrt{\chi(\Omega^*)}} + f_{\delta}(t); \\
			\\
				\ddot{\theta} = \mbox{ }-\left(\dfrac{\delta}{\tau_{\dot{\theta}}\delta_0} + \dfrac{\delta^{1/2}}{2\tau_{\delta} \delta_0^{1/2} \sqrt{\chi(\Omega^*)}} \right)\dot{\theta} + \dfrac{\delta}{\tau_{\dot{\theta}}\delta_0}\Omega + f_{\dot{\theta}}(t).
		\end{cases}
		\label{eq:model_stoch}
\end{equation}
Here, $f_{\dot{\theta}}(t)$ and $f_{\delta}(t)$ are stochastic terms that represent noise with, respectively, diffusivity $D_{\dot{\theta}}$ and $D_{\delta}$.
Thus the stochastic behaviour of the LSC is described by diffusive motions in potential wells whose shape is determined by the deterministic terms in equation~(\ref{eq:model_stoch}), which change periodically in response to the applied modulations. The potential functions are given by $V(\dot{\theta}) = -\int \ddot{\theta} \partial \dot{\theta}$ and $V(\delta) = -\int \dot{\delta} \partial \delta$.

However, in viewing that in the present study the applied modulation period is typically much longer than the characteristic timescale of the flow, dictated by the LSC turnover time ($2\pi/\omega > \mathcal{T}$), we can additionally make the simplified assumption that the diffusion of both $\dot{\theta}(t)$ and $\delta(t)$ is constrained in potential wells $V(\dot{\theta})$ and $V(\delta)$ that vary adiabatically in between different modulation phases. For a given phase $\Phi$, therefore, $V(\dot{\theta})$ and $V(\delta)$ are then assumed to be stationary with their control parameters given by their phase-average values. The governing equations then become
\begin{equation}
		\begin{cases} 
		\dot{\delta} = \dfrac{\delta}{\tau_{\delta}} - \dfrac{\delta^{3/2}}{\tau_{\delta} \langle \delta \rangle^{1/2}_{\Phi}} + f_{\delta}(t); \\
			\\
			\ddot{\theta} = \mbox{ }-\left(\dfrac{\langle \delta \rangle_{\Phi}}{\tau_{\dot{\theta}}\delta_0} + \dfrac{1}{2\tau_{\delta}} \right)\dot{\theta} + \dfrac{\langle \delta \rangle_{\Phi}}{\tau_{\dot{\theta}}\delta_0}\langle\Omega\rangle_{\Phi} + f_{\dot{\theta}}(t).
		\end{cases}
		\label{eq:model_stoch_adiab}
\end{equation}
Here $\langle \delta \rangle_{\Phi} = \delta_0\langle \lambda^2(\Omega)\rangle_{\Phi}/\lambda^2(0)$ is the time-average of $\delta$ during the phase $\Phi$ (effectively merging the factors~$\delta_0^{1/2}$ and~$\sqrt{\chi(\Omega^*)}$ into a single value for each phase) and $\langle\Omega\rangle_{\Phi}$ is the average value of $\Omega$ during this phase. We have simplified the $\dot{\theta}$-equation consistently with the adiabatic approach by using the additional approximation $\delta/\langle \delta \rangle_{\Phi} \approx 1$ within each separate phase $\Phi$. 
This approximation is valid since the relaxation time scale $\delta$, given by $\tau_{\delta}$, is much larger than that of $\dot{\theta}$. Variation of $\dot{\theta}$ is thus typically much faster than that of $\delta$. Thus, we take the time-dependent variable $\delta(t)$ to be its phase-average value for the $\ddot{\theta}$-equation.

In this adiabatic approximation, the statistic behaviour of $\dot{\theta}(t)$ and $\delta(t)$ is not dependent on previous phases, and can be evaluated separately for each phase. It is determined by the strength of the stochastic driving terms $f_{\dot{\theta}}(t)$ and $f_{\delta}(t)$, and the two potentials functions, respectively:

\begin{equation}
V(\delta) = -\frac{1}{2}\frac{\delta^2}{\tau_{\delta}} + \frac{2}{5}\frac{\delta^{5/2}}{\tau_{\delta}\langle \delta \rangle^{1/2}_{\Phi}},
\label{eq:potential_delta_adiab}
\end{equation} 
and

\begin{equation}
V(\dot{\theta}) = \frac{\dot{\theta}^2}{2}\left(\frac{\langle \delta \rangle_{\Phi}}{\tau_{\dot{\theta}}\delta_0} + \frac{1}{2\tau_{\delta}}\right) - \frac{\langle \delta\rangle_{\Phi}\langle\Omega\rangle_{\Phi}}{\tau_{\dot{\theta}}\delta_0}\dot{\theta}.
\label{eq:potential_theta_adiab}
\end{equation}

This adiabatic approach can be useful in describing phase-specific characteristics of $\dot{\theta}(t)$ and $\delta(t)$, as we will demonstrate in the following sections.

\subsection{Cessation frequency}

Here, we discuss a theoretical approach to model the shape of the modulated cessation frequency curves, displayed in Fig.~\ref{fig:Figure_14}.
We follow the approach outlined in Assaf et al.\ \cite{AAG11}, which uses the potential function of the thermal LSC amplitude to estimate the frequency of cessations. To start with, we use the potential function $V(\delta)$ from equation~(\ref{eq:potential_delta_adiab}) resulting from the adiabatic approximation. The average time $T_{*}$ it takes the thermal amplitude to reach a certain low value $\delta_* \ll \delta_0$ is now given by
\begin{equation}
T_{*}(\delta_*) = \sqrt{2\pi}\frac{\sqrt{\tau_{\delta} D_{\delta}}}{|V'(\delta_*)|} \mbox{ } e^{\frac{2}{D_{\delta}}\bigl[V(\delta_*) - V(\langle \delta \rangle)\bigr]},
\label{eq:potential_time}
\end{equation}
where $V(\langle \delta \rangle)$ is the potential evaluated at the mean value $\langle \delta \rangle = \langle \delta \rangle_{\Phi}$. We present the time $T(\delta_*)$ here using the dimensional quantities related to $\delta$; in \cite{AAG11}, the approach is presented in terms of dimensionless parameters related to $\xi \equiv \delta/\delta_0$. The reader may easily check that the equations given here are equivalent to those provided in \cite{AAG11}, by realizing that $D_{\xi} = \tau_{\delta}/\delta_0^2 \cdot D_{\delta}$, $V(\xi) = \tau_{\delta}/\delta_0^2 \cdot V(\delta)$, and $V'(\xi) = \tau_{\delta}/\delta_0 \cdot V'(\delta)$.

Correspondingly, the frequency of cessations~$\eta$ is given by

\begin{equation}
\eta^{-1} = \frac{1}{\delta_{c}}\int_0^{\delta_{c}} T_{*}(\delta_*) \partial \delta_*,
\label{eq:cess_freq}
\end{equation}
where $\delta_{c}$ is the amplitude threshold below which a cessation is defined to occur (for which we use the same criterion for~$\delta_c$ as applied to our experimental analysis).
Combining equations~(\ref{eq:potential_time}) and~(\ref{eq:cess_freq}) gives

\begin{equation}
\eta^{-1} = \sqrt{2\pi}\frac{\sqrt{\tau_{\delta} D_{\delta}}}{\delta_{c}} \int_0^{\delta_{c}} \frac{1}{|V'(\delta_*)|}  e^{\frac{2}{D_{\delta}}\bigl[V(\delta_*) - V(\langle \delta \rangle)\bigr]} \partial \delta_*.
\label{eq:master_cess}
\end{equation} 
In the case of modulated rotation rates, as seen in to our experimental results, $D_{\delta}$ becomes periodically modulated as well, $D_{\delta} = D_{\delta}(\Phi)$. This is clear from Fig.~\ref{fig:Figure_13}(a) - when $\delta(t)$ is in its minimum phase $\Phi_{min}$, for example, it is constrained by the requirement that $\delta \geq 0$. The farther away $\delta$ is modulated towards high values, the less it is influenced by this constraint. Thus, the diffusivity will reach a minimum in the minimum phase.

\begin{figure}
\centering
\includegraphics[width=0.9\textwidth]{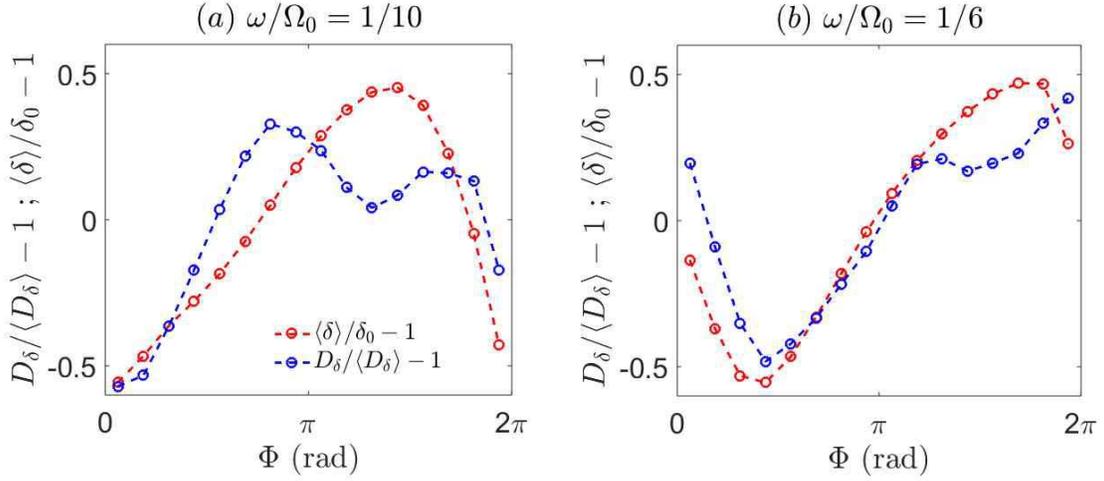}

\caption{Experimental data for the diffusivity $D_{\delta}(\Phi)$ (determined using equation~\ref{eq:msqdisp}), alongside the average value of $\delta(t)$ in each phase $\Phi$ (both divided by their mean and subtracted by one) for $\omega/\Omega_0 =$ (a) 1/10; (b) 1/6. Both variables show, roughly, the same overall development and a similar normalized amplitude.}
\label{fig:Figure_18}
\end{figure}

We prove this inference by calculating the diffusivity of $\delta$ as a function of the phase, $D_{\delta}(\Phi)$, from the experimental data as follows. First, we calculate the mean-square displacement $\psi$ of $\delta(t)$ for each phase $\Phi$:
\begin{equation}
\psi(\tau)|_{\Phi = \Phi_n} = \left\langle \left( \delta(t + \tau)|_{\Phi = \Phi_n} - \delta(t)|_{\Phi = \Phi_n} \right)^2 \right\rangle_t -  \left\langle \left( \delta(t + \tau)|_{\Phi = \Phi_n} - \delta(t)|_{\Phi = \Phi_n} \right) \right\rangle_t^2.
\end{equation}
Here, the subscript $(...)|_{\Phi = \Phi_n}$ means that the variable in question is evaluated only within a certain phase $\Phi_n$ out of $N$ total phases within one period in the range $0 \leq t \leq 2\pi/\omega$.
The diffusivity of $\delta(t)|_{\Phi = \Phi_n}$ is then given by a linear fit of the form
\begin{equation}
\psi(\tau)|_{\Phi = \Phi_n} \sim D_{\delta}(\Phi_n) \tau
\label{eq:msqdisp}
\end{equation}
in the range $0 \leq \tau \leq 30$~s, where such a fit is typically possible.
In Fig.~\ref{fig:Figure_18}, we show the normalized values of $D_{\delta}(\Phi)$ obtained in this way, along with the values of the temporal mean $\langle \delta \rangle_{\Phi}$ in each phase, for two different $\omega/\Omega_0$. It can be seen that, indeed, both shapes roughly follow the same development, with minima globally occurring in similar phases $\Phi$, and sharp increases in $D_{\delta}$ corresponding to sharp increases in $\langle \delta \rangle_{\Phi}$. Indeed, even the amplitude-to-mean ratio of both variables is roughly the same.

We now proceed by modeling the phase dependence of $\eta$ by inserting the experimental values of $D_{\delta}(\Phi)$ and $\langle \delta \rangle$ into equation~(\ref{eq:master_cess}). The value $D_{\delta}(\Phi)$ enters the equation both in the prefactor as well as in the exponent of the integrand. The value $\langle \delta \rangle$ is used to evaluate the term $V(\langle \delta \rangle)$ in the exponent.
Results from equation~(\ref{eq:master_cess}) for $\omega/\Omega_0 = [1/10, 1/6]$ are plotted in Fig.~\ref{fig:Figure_14}(a)-(b) along with the experimental results. It can be seen that the modeled shapes of the cessation frequency are close to what has been experimentally measured, with a distinct peak in cessations occurring in or close to the phase $\Phi_{min}$. Furthermore, the experiment and model results both exhibit a broader peak for the higher modulation rate $\omega/\Omega_0 = 1/6$. Thus, it appears that our approach strengthens the theory of Assaf et al.\ \cite{AAG11, AAG12} by replicating closely an experimentally-observed temporally modulated cessation frequency.

We note here that the shape of $\eta(\Phi)$ according to equation~(\ref{eq:master_cess}) is found to be insensitive to the value chosen for $\delta_c$, but the absolute values of $\eta(\Phi)$ are not. Using $\delta_c = 0.10\delta_0$ as in the experiments, absolute values from the model are somewhat higher than experimentally measured; for instance, for $\omega/\Omega_0 = 1/6$, the maximum in $\eta$ is then~$9\cdot 10^{-3}$~s$^{-1}$, as compared to $3.5\cdot 10^{-3}$~s$^{-1}$ from the experimental data. In order to obtain results that match closely with the experimental values, a value $\delta_c \approx 0.05\delta_0$ would have to be used in the model. This difference in absolute values of cessation frequency between model and experiments is likely down to two reasons. Firstly, $\eta$ is extremely sensitive to the exponential term given by both $D_{\delta}$ and $\langle \delta \rangle$, but we use here their arithmetic mean in each phase, as an approximation. This may cause part of the differences in the magnitude of $\eta$. Secondly, and more importantly, in equation~(\ref{eq:cess_freq}), it is assumed that the PDF $P(\delta_{*})$, representing the fraction of cessation events in which the minimum of $\delta$ is $\delta_{*}$, taken across all cessation events (within a phase), is a constant. However, experimental data shows that $P(\delta_{*})$ decreases when $\delta_{*}$ decreases, so we should in theory integrate equation~(\ref{eq:cess_freq}) over $P(\delta_{*})\partial(\delta_{*})$ to obtain $\eta$. The approximation treating $P(\delta_{*})$ as a constant here has overestimated $\eta$.

\subsection{Probability distribution of $\delta$}

Here, we discuss an approach allowing to model the shape of the probability distribution of $\delta$. The probability distribution is given by
\begin{equation}
P(\delta) = C_{\delta} e^{-2V(\delta)/D_{\delta}},
\label{eq:pdelta_basis}
\end{equation}
where $C_{\delta}$ is the appropriate normalization constant. We have seen that the shape of the PDFs depends on the phase $\Phi$ (see Fig.~\ref{fig:Figure_15}). Thus, it is important to take this dependency along here, which we do using the ``adiabatic'' approximation from equation~(\ref{eq:potential_delta_adiab}).

We fitted the function $P(\delta)$ to the experimentally obtained PDFs by varying the parameters $D_{\delta}$ and $\langle \delta \rangle_{\Phi}$ within realistic ranges. (The value of $C_{\delta}$ always results from the requirement that the area under the PDF be unity, and is not further relevant.)
Upon fitting $P_{\delta}$ to the experimental PDFs corresponding to $\Phi = \Phi_{max}$ in Fig.~\ref{fig:Figure_15}, we found that a least-squares fitting procedure resulted in best-fit parameters corresponding to $D_{\delta} = 6.4\cdot 10^{-5}$~K$^2$/s and $\langle \delta \rangle_{\Phi} = 0.33$~K.  Clearly, these parameters are quite comparable to experimental values for $\Phi = \Phi_{max}$, with $D_{\delta} \sim 3 \cdot 10^{-5}$~K$^2$/s and $\langle \delta \rangle \approx 0.3$~K during the maximum phase (cf.\ Fig.~\ref{fig:Figure_13} for the latter). 
The same procedure applied to $\Phi = \Phi_{min}$ resulted in $D_{\delta} = 5.8\cdot 10^{-5}$~K$^2$/s and $\langle \delta \rangle_{\Phi} = 0.22$~K. These values diverge more strongly from experimental results, with $\langle \delta \rangle \approx 0.10$~K during the minimum phase for both $\omega/\Omega_0 = 1/10, 1/6$. 

The curves $P(\delta)$ corresponding to these parameters are shown as smooth lines in Fig.~\ref{fig:Figure_15}. It can be seen that of the PDF is represented extremely well for $\Phi_{max}$, but slightly less so for $\Phi_{min}$.
We assume that one chief reason this approach fails to work well for the minimum phase is that the mathematical model in this form cannot account for the steep drop-off of the PDF that happens in the vicinity of zero, as the increased occurrence of cessations is not explicitly modeled in equation~(\ref{eq:potential_delta_adiab}). While the value of $\delta$ can never be lower than zero, this only presents a real inaccuracy in the model in the minimum phase, when the most likely values of $\delta$ are concentrated much closer to zero than during other phases.

\subsection{Probability distribution of $\dot{\theta}$}
\label{subsec:pdf_dtheta}

Lastly, we can use our modeling approach to replicate some of the statistical features of the probability distributions of $\dot{\theta}$, experimental results of which are given in Fig.~\ref{fig:Figure_16}. We demonstrate two approaches in this section: (1) Assuming the conditional PDF $P(\dot{\theta} | \delta)$ equilibrates much faster than the time scale of $\tau_{\delta}$, and then following the calculations as those in~\cite{AAG11}; and (2) a simplification of this approach using the adiabatic approximation.

Under approach~(1), we calculate the probability distribution of $\dot{\theta}$ in the model as follows:

\begin{equation}
P(\dot{\theta}) = \int_0^{\infty} P(\dot{\theta} | \delta) P(\delta) \partial \delta.
\label{eq:pdottheta}
\end{equation}
Here, the steady-state conditional PDF $P(\dot{\theta} | \delta)$ is given by

\begin{eqnarray*}
P(\dot{\theta} | \delta) &=& C_{\dot{\theta}} e^{-2V(\dot{\theta})/D_{\dot{\theta}}} \\
&=& C_{\dot{\theta}} e^{-\left(\left[\frac{\delta}{\tau_{\dot{\theta}}\delta_0} + \frac{\delta^{1/2}}{2\tau_{\delta}\delta_0^{1/2}\sqrt{\chi(\Omega^*)}}\right] \dot{\theta}^2 - 2\frac{\delta\Omega\dot{\theta}}{\tau_{\dot{\theta}}\delta_0}\right)/D_{\dot{\theta}}}
\end{eqnarray*}
(where $C_{\dot{\theta}}$ is a normalization constant, and $D_{\dot{\theta}}$ is the diffusivity of $\dot{\theta}$; we use the value $D_{\dot{\theta}} \approx 8 \cdot 10^{-5}$~rad$^2$s$^{-3}$ as estimated from our experiments), and the PDF $P(\delta)$ is given by

\begin{eqnarray*}
P(\delta) &=& C_{\delta} e^{-2V(\delta)/D_{\delta}} \\
&=& C_{\delta} e^{-\left(-\frac{\delta^2}{\tau_{\delta}} + \frac{4}{5}\frac{\delta^{5/2}}{\tau_{\delta} \delta_0^{1/2} \sqrt{\chi(\Omega^*)}}\right)/D_{\delta}(\Phi)}.
\end{eqnarray*}
In equation~(\ref{eq:pdottheta}), therefore, the dependence on $\Phi$ appears implicitly in $D_{\delta}$ ($D_{\dot{\theta}}$ has little dependence on $\Phi$, as $\dot{\theta}$ is not constrained by the requirement to be positive, unlike $\delta$) as well as in the Coriolis term $\sim \Omega$ in the exponent of $P(\dot{\theta} | \delta)$ and in the factor $\chi(\Omega^*)$.

\begin{figure}
\centering
\includegraphics[width=0.95\textwidth]{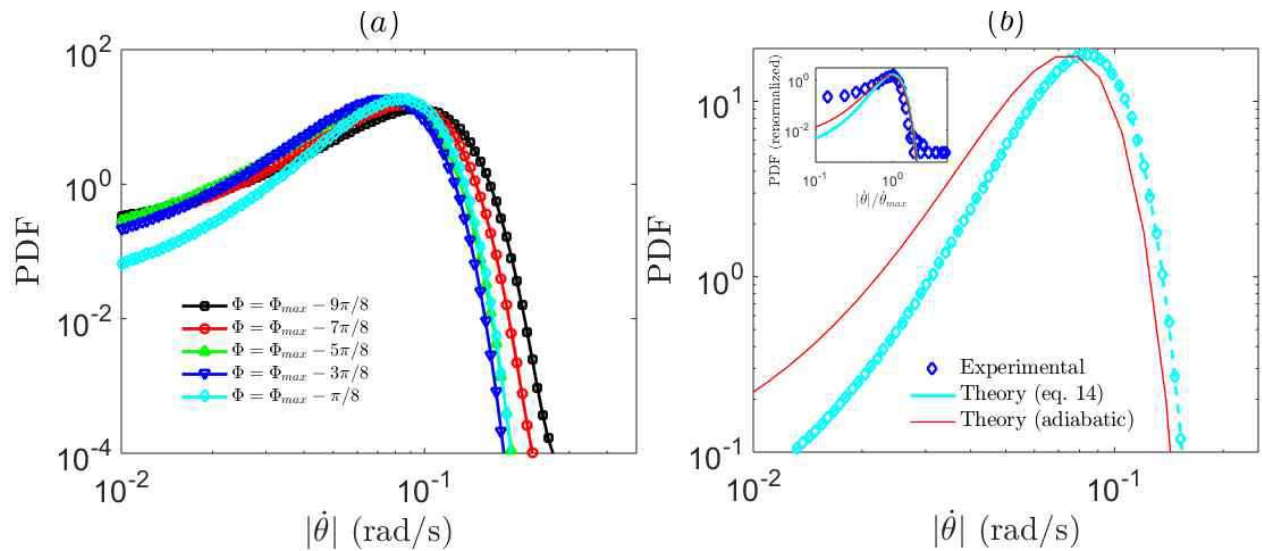}
\caption{(a) Results from equation~(\ref{eq:pdottheta}) for the PDFs of $|\dot{\theta}|$ in various phases $\Phi$, for model parameters corresponding to the experiment with $\omega/\Omega_0 = 1/10$. (b) The experimental results for $\Phi = \Phi_{max} - \pi/8$ compared to model results from two approaches: (1) from equation~(\ref{eq:pdottheta}); (2) from using the adiabatic simplification $P(\dot{\theta}) = e^{-2V(\dot{\theta})/D_{\dot{\theta}}}$ with $V(\dot{\theta})$ as in equation~(\ref{eq:potential_theta_adiab}), for $\omega/\Omega_0 = 1/10$ and $\Phi = \Phi_{max} - \pi/8$. Inset: when normalizing the horizontal axis by the value $\dot{\theta}_{max}$, where the maximum probability occurs, and renormalizing each PDF such that its integral is unity, experimental and model results roughly collapse for data sets where $\Phi$ is near $\Phi_{max}$.}
\label{fig:Figure_19}
\end{figure}

Calculating results from equation~(\ref{eq:pdottheta}) using the above equations gives the shapes of $P(\dot{\theta})$ for each value of $\Phi$. A number of these shapes for different phases, for model parameters corresponding to the experiment with $\omega/\Omega_0 = 1/10$, are displayed in Fig.~\ref{fig:Figure_19}(a). We can easily compare the fall-off slope of these shapes to those experimentally observed in Fig.~\ref{fig:Figure_16}(a); the results are given in Fig.~\ref{fig:Figure_16}(b) (in each case, slopes were calculated roughly in the decade beyond the value of $\dot{\theta}$ where the PDF showed a maximum). It is seen that the dependence of this fall-off shows a peak, like the experimental data, and the synchronization with the experimental results is also clear, indicating again the adequateness of the phase-dependent terms in the full dynamical model in equation~(\ref{eq:model}). The peak is, however, much more pronounced in the experimental data than in the model results. The reason is that the fall-off slope is, among other things, a proxy for how many cessations happen in a certain phase (a very high slope indicates the absence of cessations), and cessations are not modeled explicitly in equation~(\ref{eq:model_stoch}), but rather implicitly through the fact that, when $\delta \ll \delta_0$, the stochastic term becomes dominant. This does, however, not model the actual jump magnitude $|\Delta \theta_c|$ as displayed in Fig.~\ref{fig:Figure_17}, resulting in a smaller range of $\epsilon$ than derived from experiments.

Previous work \cite{AAG11} explained how this theoretical approach can be used to predict a slope of $\epsilon = -4$ in the tails of $P(\dot{\theta})$ when $\delta \ll \delta_0$, representing the LSC undergoing cessations. Indeed, we see in Fig.~\ref{fig:Figure_16}(b) that the minimum values found for $\epsilon$ (i.e.\ in the phases where cessations are most likely) are quite close to this: $-3.3$ for the experimental data and $-4.2$ for the results from the model (Assaf et al.\ \cite{AAG11} found -4.3 for nonrotating RB convection in a wide range of $Ra$).
We note also, however, that the model approach does not replicate the experimentally observed absence of a sharp peak probability for phases where the fall-off slope is close to~$-4$; instead, the model predicts a clear peak probability for all~$\Phi$ whose position changes only minimally with~$\Phi$. 

This is a relatively involved calculation, requiring experimental data on the diffusivity of $\delta$ (likely subject to uncertainty) as well as a separate calculation of the potential of $\delta$ before that of $\dot{\theta}$ can be inferred.
We show here that approach (2), using the adiabatic potential $V(\dot{\theta})$ from equation~(\ref{eq:potential_theta_adiab}), can give similar results to replicate the shapes of the probability distribution of $\dot{\theta}$. The fact that we use this adiabatic assumption, in turn, justifies using the assumption $P(\dot{\theta}) \approx P(\dot{\theta}|\delta)$ \textit{within} each phase $\Phi_n$. We have used equation~(\ref{eq:potential_theta_adiab}) to calculate the PDF $P(\dot{\theta})$ for the curve corresponding to $\Phi = \Phi_{max} - \pi/8$ (close to the maximum phase; see Fig.~\ref{fig:Figure_16}(a)); the result is given in Fig.~\ref{fig:Figure_19}(b). With the parameters $\langle \delta \rangle_{\Phi} \approx 0.3$~K (close to the experimental value) and $D_{\dot{\theta}} \approx 8.0\cdot 10^{-5}$~rad$^2$s$^{-3}$, the shape of the curve from the full equation can be closely approximated. The slope of the tail of this PDF is again very close to what the experimental data suggest. 

Both theoretical approaches slightly overestimate the value of $\dot{\theta}$ where the PDF exhibits a peak (compare Fig.~\ref{fig:Figure_16}(a) and~\ref{fig:Figure_19}), although experimental and theoretical values of this $\dot{\theta}$ are all of order~0.01~rad/s. As discussed for the results in Fig.~\ref{fig:Figure_9}(d)-(e), the discrepancy could again be due to an underestimation of the LSC inertia. However, if we plot the PDFs as a function of the normalized variable $|\dot{\theta}|/\dot{\theta}_{max}$, where $\dot{\theta}_{max}$ is the value with maximum probability, and accordingly renormalize the PDF such that its integral is equal to unity, the theoretical and experimental curves collapse extremely well onto each other, as can be seen in the inset to Fig.~\ref{fig:Figure_19}(b). The adiabatic approach is, in fact, even closer to the experimental curve than the result from the full equation~(\ref{eq:pdottheta}).

In conclusion, both approaches (1) and (2) give very similar results and work well in replicating the fall-off slope of the PDFs of $\dot{\theta}$ especially near $\Phi_{max}$. Approach (2) requires fewer estimations of phase-dependent parameters from experimental data and is not coupled to the equation for the PDF of $\delta$, a quantity to which $\dot{\theta}$ is dynamically coupled according to equation~(\ref{eq:model_stoch}); thus, it is a much simpler approach to obtain very similar results. In phases near $\Phi_{min}$, the experimentally observed shape of the probability is, however, less well replicated by these approaches, so further refinements to the theory are needed to improve our understanding of the phase-dependence of the PDFs.

\section{Results for heat transport}

Our results on rotating Rayleigh-B\'enard convection discussed up to this point have focused on the dynamics of the large-scale circulation. All the responses to the modulation of the frame of reference's rotation rate discussed so far are related to the dynamical response of the large-scale circulation orientation and strength.

However, this could obscure other responses to the modulated rotation rate which may be present in the background and unrelated to the LSC, that may still have ramifications for instance for overall heat transport. In the results discussed in this section, we have tried to take the LSC out of the equation by moving to a different parameter range, and focus on the response of the turbulent fluid motions when there is no LSC to influence. We first discuss how the parameter range and the modulation range were chosen based on results for constant rotation, before moving on to discuss the results of modulated-rotation experiments.

\subsection{Results for constant rotation}

We note that, when the rotating speed $\Omega$ increases beyond $1/Ro \gtrsim 0.8$, we do not detect any clear signature of an LSC from the side-wall temperature signals (see also \cite{ZA10}). Indeed, when $1/Ro$ increases, the increasing Coriolis force alters the flow field from one turbulent state, dominated by the LSC flow, to another turbulent state in which local thermal plumes are organized into long columnar vortices. These vortical plumes are coherent thermal structures that give rise to the enhancement of the global heat transport, known as the Ekman-pumping effect~\cite{ZSCVLA09}. 

\begin{figure}
\centering
\includegraphics[width=0.6\textwidth]{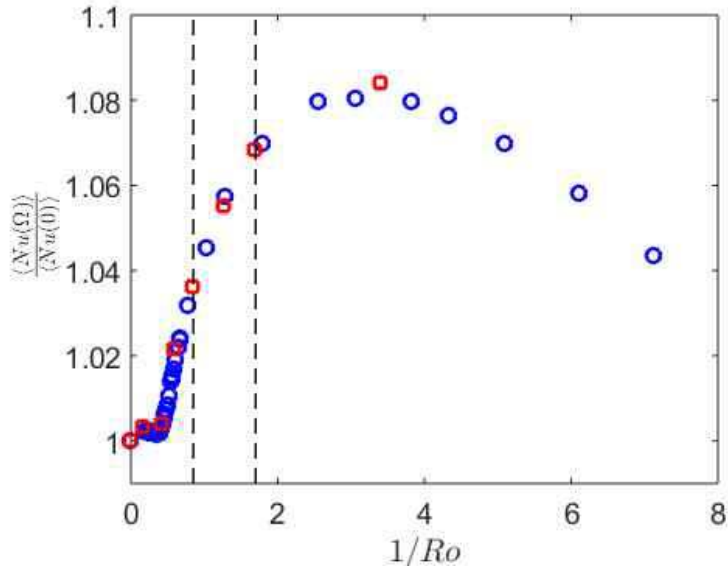}
\caption{The dependency of $Nu$ on changing (constant) rotation rate. Plotted are the time-averages of $Nu$ normalized by the time-average at zero rotation against $1/Ro$. Blue circles: experimental data from~\cite{ZA10} with $Ra = 2.25\times 10^9$; red squares: the present work with $Ra = 2.1\times 10^9$. The two dashed lines indicate the range $0.85 < 1/Ro < 1.70$ in which we performed modulated-rotation experiments.}
\label{fig:Figure_20}
\end{figure}

In Fig.~\ref{fig:Figure_20}, we show results for the Nusselt number $Nu$ (normalized by its value at zero rotation) as a function of $1/Ro$ for $Ra = 2.1\times10^9$, as well as results from~\cite{ZA10} at the similar value $Ra = 2.25\times10^9$. It can be observed that an enhancement of the global heat transport with increasing rotation rates occurs within the range $0.4 < 1/Ro < 2.5$. Furthermore, it has been found~\cite{ZA10} that the enhancement of $Nu$ with $1/Ro$ decreases as $Ra$ becomes larger.

\begin{figure}[t!]
\centering
\includegraphics[width=0.8\textwidth]{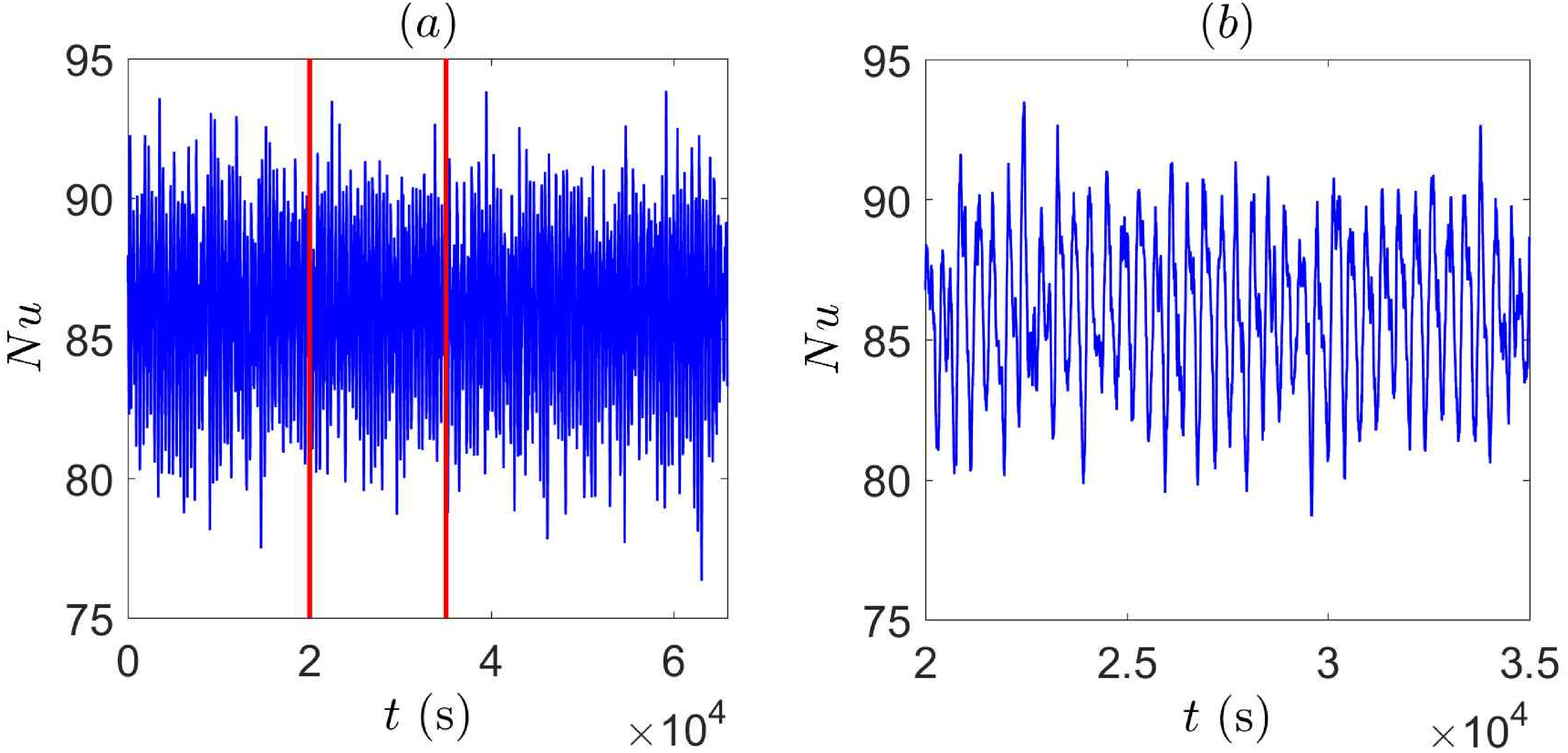}
\caption{The time trace of $Nu$ for a $\omega/\Omega_0 = 1/10$ heat transfer experiment (left). A close-up of the response of $Nu$ between the two red lines is given in the right plot, in which the oscillations are clearly visible.}
\label{fig:Figure_21}

\vspace{1.0cm}
\centering
\includegraphics[width=0.8\textwidth]{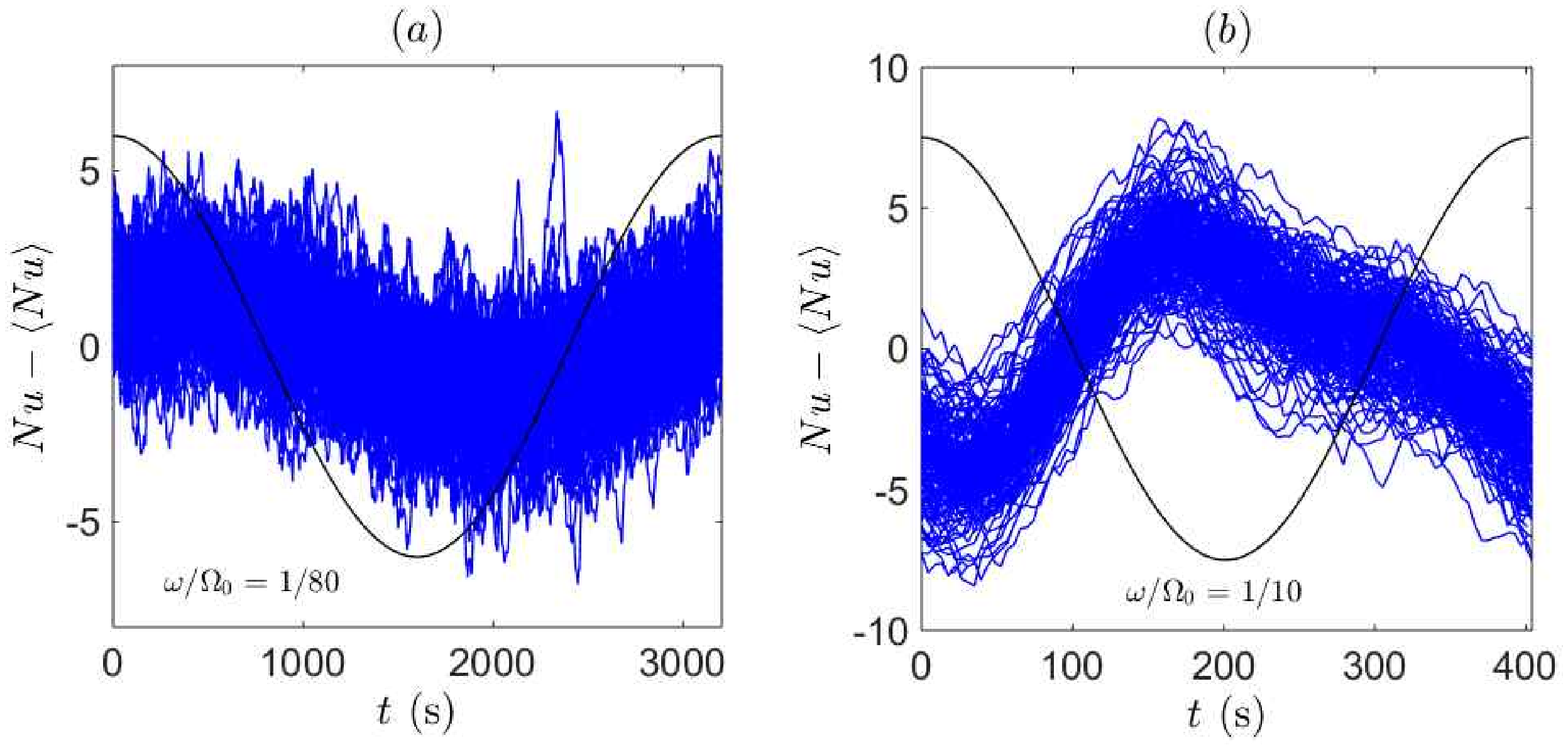}
\caption{The ensemble response of $Nu$ for two experiments at different $\omega/\Omega_0$. The maxima in $\Omega(t)$ are timed at $t = 0$, which is plotted in arbitrary units as solid black line.}
\label{fig:Figure_22}
\vspace{1.0cm}
\includegraphics[width=0.8\textwidth]{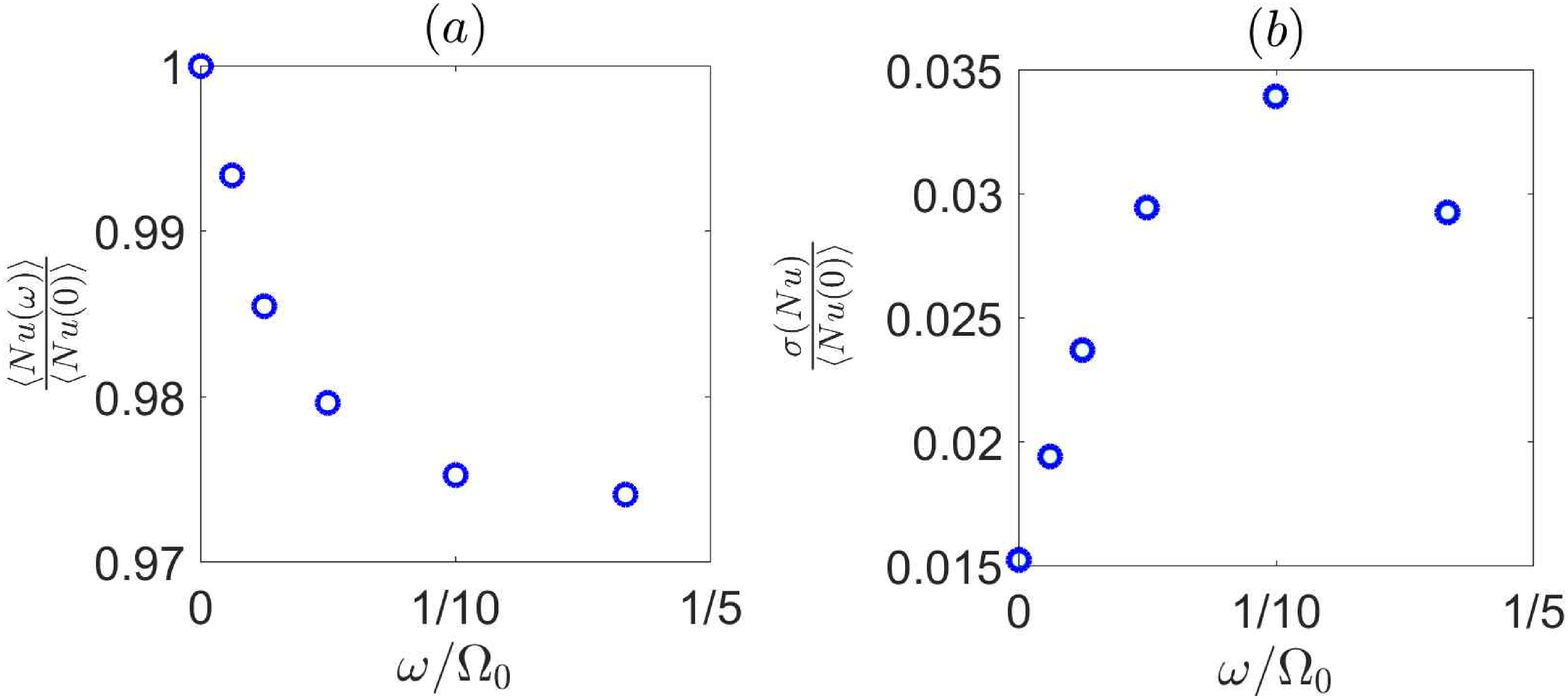}
\caption{The mean (a) and standard deviation (b) of $Nu$, normalized by the mean of $Nu$ at zero modulation, versus $\omega/\Omega_0$.}
\label{fig:Figure_23}
\end{figure}

\begin{figure}[t!]
\centering
\includegraphics[width=0.7\textwidth]{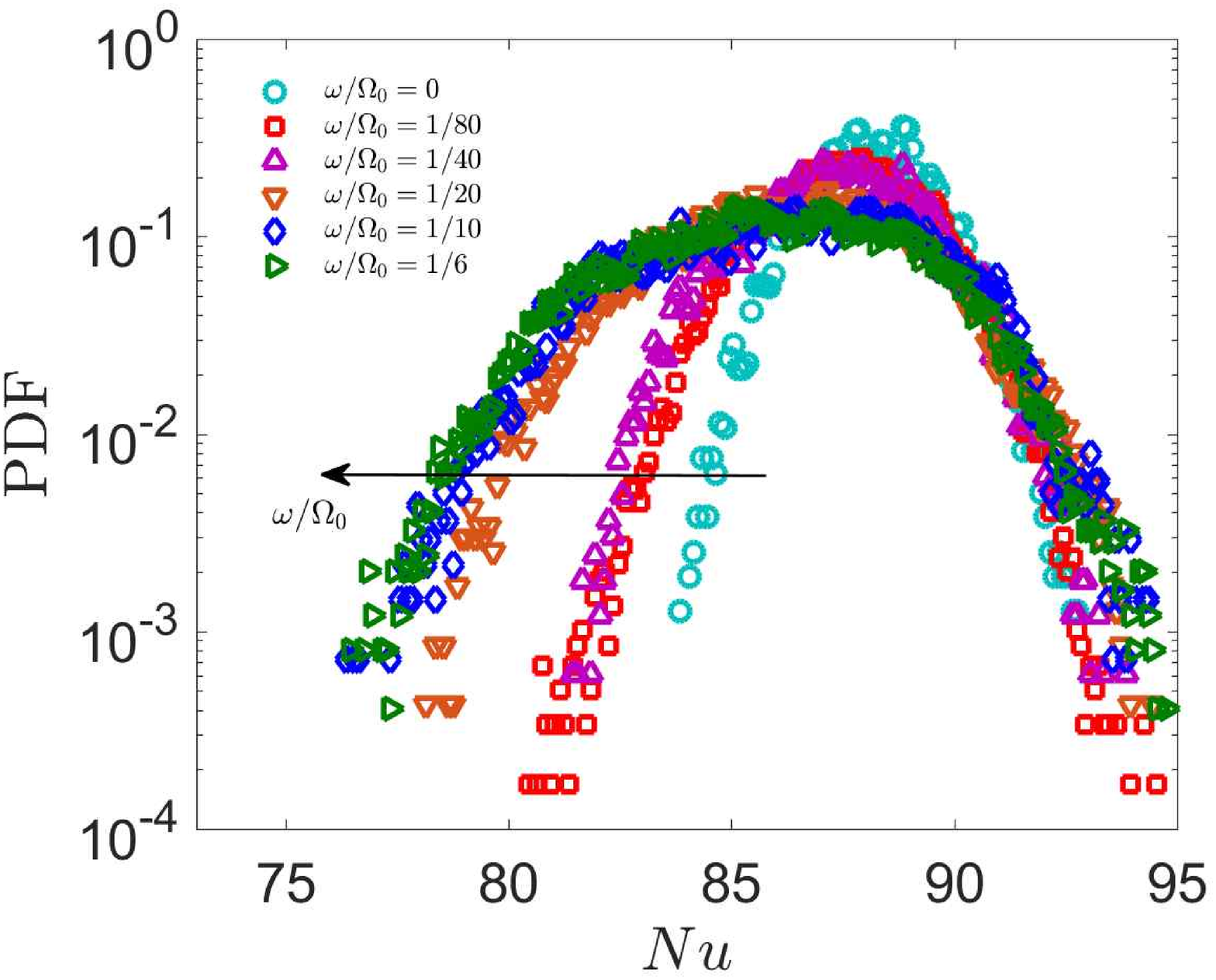}
\caption{The PDFs of $Nu$ for different $\omega/\Omega_0$. The right (high-$Nu$) tails are seen to be near-invariant with $\omega/\Omega_0$. The left (low-$Nu$) tails, on the other hand, seem to move to lower and lower values as $\omega/\Omega_0$ is increased.}
\label{fig:Figure_24}
\vspace{1.0cm}
\centering
\includegraphics[width=0.5\textwidth]{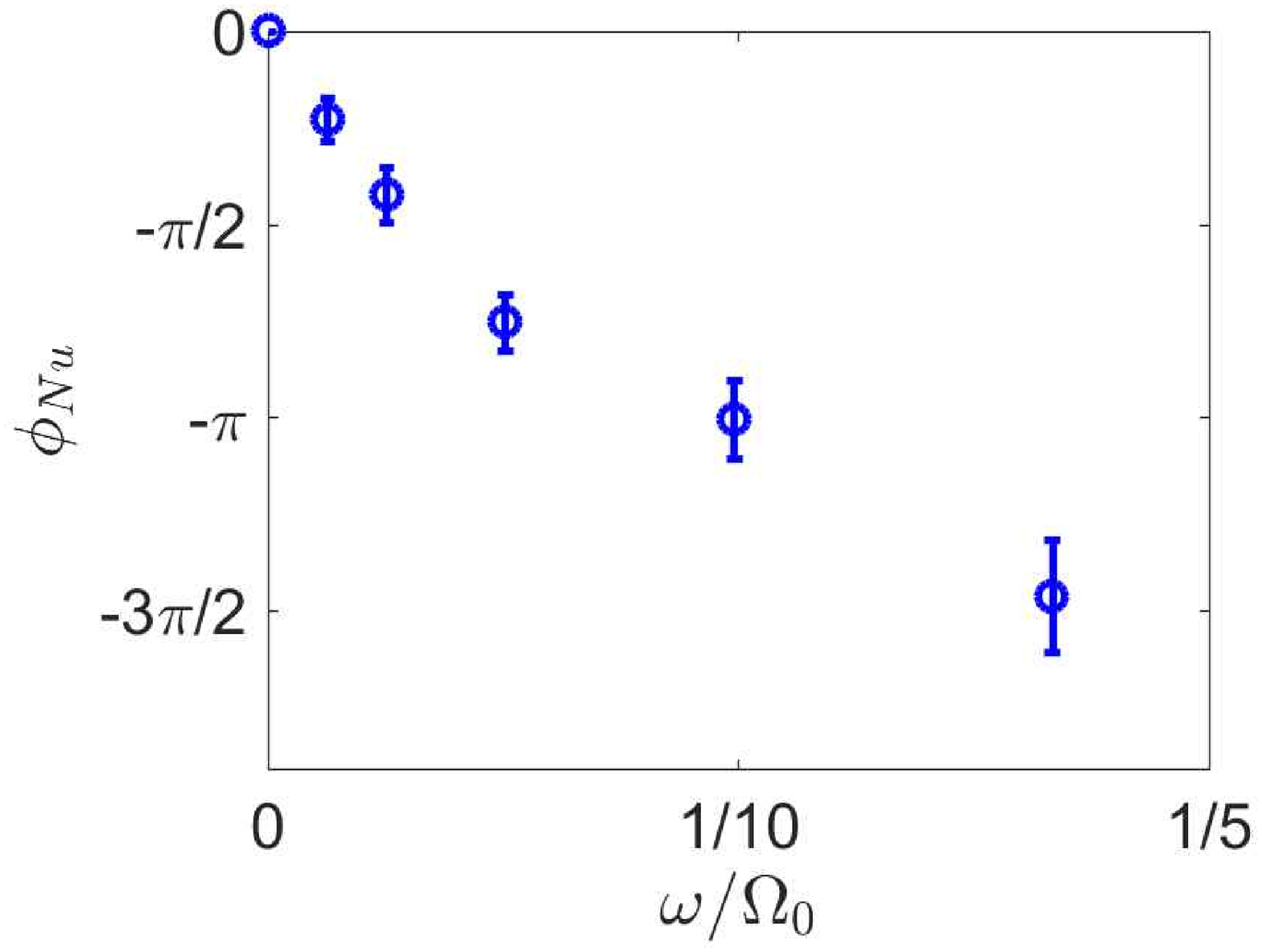}
\caption{The phase lag of the oscillations in $Nu$ versus $\omega/\Omega_0$.}
\label{fig:Figure_25}
\end{figure}

\subsection{Results for modulated rotation}

Based on these experimental data, we decided to perform modulated-rotation experiments at $Ra = 2.1\times 10^9$ in which we set $1/Ro = 1.27$ as the mean inverse Rossby number, with $0.85 < 1/Ro < 1.70$ as the range of libration. The boundaries of this range are indicated by vertical lines in Fig.~\ref{fig:Figure_20}. 
This range corresponded to a mean rotation rate of $\Omega_0 = 0.157$~rad/s and modulation amplitude $\beta = 0.33$. The choice was based, firstly, on the fact that a strong ``adiabatic" response of $Nu$ to changing $1/Ro$ is to be expected in this range, as seen in Fig.~\ref{fig:Figure_20}, and secondly on the fact that the rotary table used for our experiments could not run in a modulation mode at higher rotation rates than the corresponding $\Omega_0(1+\beta) = 0.209$~rad/s.

In our experiments, we find that the Nusselt number shows a periodic response at the frequency of modulation. We show an example of a time trace of~$Nu$ in Fig.~\ref{fig:Figure_21}, from an experiment with~$\omega/\Omega_0 = 1/10$. In Fig.~\ref{fig:Figure_21}(b), the oscillatory character of $Nu$ can be clearly seen. We can construct ensemble oscillations of $Nu$ from such time traces, using the same methodology as applied earlier for ensembles of $\dot{\theta}$ and $\delta$ (described in section~III). In Fig.~\ref{fig:Figure_22}, we plot the ensembles of modulated $Nu$ for two example values of~$\omega/\Omega_0$. (The maxima of $\Omega(t)$ occur at $t = 0$ in these plots.) We see that, at the (very low) value~$\omega/\Omega_0 = 1/80$, short-timescale deviations affect the periodic response much more strongly than at~$\omega/\Omega_0 = 1/10$. 

We plot the mean value $\langle Nu \rangle$ and the standard deviation $\sigma(Nu)$ (both taken, in each case, across an entire experiment minus transient periods in the beginning) against $\omega/\Omega_0$ in Fig.~\ref{fig:Figure_23}. Since very strong deviations from the ensemble response are present for low $\omega/\Omega_0$, and we have no criterion for identifying them (as we did for cessations in the discussion of LSC dynamics), we cannot determine the amplitude of oscillations in $Nu$ from the peak-to-peak amplitude of the response. Instead, we use the standard deviation $\sigma(Nu)$ as a measure of the fluctuations in $Nu$. (Alternatively, one could approximate the amplitude from the mean values of $Nu$ during different phases of one modulation cycle. We find that the resulting trend is identical to that of $\sigma(Nu)$ in Fig.~\ref{fig:Figure_23}, except that we cannot define the value at $\omega/\Omega_0 = 0$ in this way.)
We see that $\langle Nu \rangle$ decreases with $\omega/\Omega_0$ in the investigated range; its relative change is about 2-3\%. This change is comparable to what we expect from the trend of $Nu$ with $1/Ro$ as given in Fig.~\ref{fig:Figure_20}. 
We also see that $\sigma(Nu)$ initially increases with $\omega/\Omega_0$ in the investigated range, but seems to peak around $\omega/\Omega_0 = 1/10$.

A possible qualitative explanation of these phenomena could be as follows. As mentioned, in the investigated regime of modulation, there are no traces of a large-scale circulation in the fluid. Heat transfer is therefore accounted for by turbulent heat transfer through thermal plumes emitted from the thermal boundary layers. When constant rates of rotation are applied such that the inverse Rossby number falls into the studied range $0.85 < 1/Ro < 1.70$, these thermal plumes are organized into columnar vortices stretched out into the bulk fluid, which enhance the momentum and heat transport through the boundary layers. When modulation of the rotating velocity~$\Omega$ is applied, however, the variation in $\Omega$ produces a time-dependent Coriolis force that may disintegrate the columnar vortical plumes into interspersed thermal fluid parcels, and suppress the effect of Ekman pumping.  Thus the applied modulation decreases the overall heat transport, and the higher the modulation frequency, the stronger will be the influence of the weakening of the coherent plume structures. In that case, the decrease factor of $Nu$ (about 2-3\% decrease between $\omega/\Omega_0 = 0$ and $\omega/\Omega_0 = 1/6$) could perhaps give quantitative clues on how the strength of Ekman-pumping suppression depends on the frequency of modulation.

The observed peak in standard deviation is then explained as follows. The standard deviation of the oscillating quantity $Nu$ serves as a proxy for its amplitude, which initially increases with $\omega$, since the additional variation in how the thermal plumes are organized (due to the partial disintegration of the columnar vortices they constitute at zero modulation) increases the temporal variation in heat transfer in the fluid. However, in the limit of high~$\omega$, the modulation becomes so fast that the effect of Ekman pumping is maximally suppressed, limiting the temporal variation of $Nu$ again. There will thus be a maximum variability in heat transport at finite $\omega/\Omega_0$.

The trends of $\langle Nu \rangle$ and $\sigma(Nu)$, taken together, imply that the maximum values of heat transfer are quite insensitive to changes in the modulation amplitude, but the minimum values are not, up to at least $\omega/\Omega_0 = 1/6$ in this particular parameter range. This can be more clearly illustrated by plotting the PDFs of $Nu$ for each $\omega/\Omega_0$, as has been done in Fig.~\ref{fig:Figure_24}. The right (high-$Nu$) tails of all five PDFs overlap roughly, whereas the left tails move to ever lower values of $Nu$ with increasing $\omega/\Omega_0$.

Lastly, we have investigated the phase shift of $Nu$ with respect to $\Omega(t)$. This phase shift has been calculated using the same algorithm as used for the phase shifts $\phi_{\dot{\theta}}$ and $\phi_{\delta}$ (described in section~III and Appendix~B). In Fig.~\ref{fig:Figure_25}, we plot the phase $\phi_{Nu}$ versus $\omega/\Omega_0$. Here, $\phi_{Nu}$ has been defined as zero when $Nu$ oscillates \textit{in phase} with $\Omega(t)$, since $Nu$ increases with $1/Ro$ in the investigated range $0.85 < 1/Ro < 1.70$, and its adiabatic response is thus in phase with $\Omega(t)$ (see Fig.~\ref{fig:Figure_20}). The oscillations in $Nu$ are seen to increasingly lag the rotation of the sample as $\omega/\Omega_0$ increases. The largest lag recorded here is three-quarters of a cycle, at $\omega/\Omega_0 = 1/6$. At modulation rates faster than $\omega/\Omega_0 = 1/6$, the oscillatory signals got mostly lost in the fluctuations and could therefore not be analyzed.

\section{Conclusion and discussion}

In this paper, we have presented comprehensive experimental and modeling results on the effects of modulated external rotations on the dynamical and statistical responses of convective circulations and heat transfer in Rayleigh-B\'enard convection at $Pr = 4.38$ and $\Gamma = 1$, with $Ra \sim 10^9$. Here we summarize the results and provide recommendations for future research.

We have measured the response of the azimuthal velocity~$\dot{\theta}(t)$ and thermal amplitude~$\delta(t)$ of the large-scale circulation (LSC) under external modulated (unidirectional) rotations of the RB cell. We have found that in the limit of very slow modulation rates $\omega$, the responses of $\dot{\theta}(t)$ and $\delta(t)$ are modulated adiabatically, i.e.\ following the dependence of $\langle \dot{\theta} \rangle$ and $\langle \delta \rangle$ on modulation speed $\Omega$ without phase delay and with the same amplitude response, as should be expected. However, increasing the modulation rate $\omega$ results in a variety of dependencies. Both $\dot{\theta}(t)$ and $\delta(t)$ exhibit increasing phase delays ($\phi_{\dot{\theta}}$ and $\phi_{\delta}$) with respect to $\Omega(t)$ as $\omega$ increases; $\phi_{\dot{\theta}}$ approaches $-\pi/2$ for high $\omega$, whereas $\phi_{\delta}$ does not seem to have an asymptotic limit for high~$\omega$. The amplitude responses of $\dot{\theta}(t)$ and $\delta(t)$ (respectively, $A_{\dot{\theta}}$ and $A_{\delta}$) both approach zero for very high~$\omega$, but the former has a peak at finite $\omega$ whereas the latter decreases monotonically. Beyond a critical modulation rate~$\omega = \omega_c$, the oscillatory signals become too weak to be discernible in the noisy background of the measured time series.

We have formulated a modeling approach that is an extension of earlier work~\cite{BA06a, BA07} to include the effects of modulated rotations. This simple approach consists of Langevin-type equations for $\dot{\theta}(t)$ and $\delta(t)$, and takes into account the effects of a modulated Coriolis force as well as the dependence of momentum BL thickness on rotation rate. The model is successful in predicting each of the qualitative trends described above, including the peak in $A_{\dot{\theta}}$, which is explained as an optimal coupling between the rotation rate $\Omega(t)$ and the thermal amplitude $\delta(t)$ in the Coriolis acceleration term in the dynamical equation for $\dot{\theta}(t)$.

As described in previous studies, the occurrence of stochastic cessation/reorientation events of an LSC is sensitive to external factors such as the rotation rate. In this study, we have extensively studied the dependence of frequency of cessation events $\eta$ on the modulation rate, and identified a sharp maximum in $\eta$ at finite $\omega$. Experimental runs with $\omega$ set very close to this value allowed us to collect enough data to statistically analyze cessation events under external modulation (at zero rotation, they roughly occur only once every three hours; at this finite $\omega/\Omega_0$, they occur~10-20 times as often). We identify a very clear dependence of the probability of cessation on the specific phase ($\Phi$) within one period of modulation, with most cessation events occurring in the phase of modulation during which $\delta(t)$ reaches its minimum values anyway.

We have extended previous modeling approaches~\cite{AAG11,AAG12} in a consistent manner with our modeling of the dynamical responses of $\delta(t)$ with an adiabatic approach to estimate the frequency of cessation and its dependence on $\Phi$ numerically. Besides the $\omega$-dependency present in the dynamical equation for $\delta(t)$, the cessation model includes an estimation of the effective diffusivity of $\delta(t)$ during different phases of a modulation cycle. The model reproduces well the experimentally found dependency of $\eta$ on $\Phi$, with the timing of the maximum in $\eta$ and the increase in the peak width of $\eta$ with increasing~$\omega$ being replicated closely.

We furthermore find that the shapes of the probability distributions of $\dot{\theta}(t)$ and $\delta(t)$ change significantly with $\Phi$, with their standard deviations and skewnesses being strongly dependent on, among other things, the probability of cessations occurring in each phase $\Phi$. We have extended modeling approaches for the PDFs of $\dot{\theta}(t)$ and $\delta(t)$~\cite{AAG11,AAG12} using the same adiabatic approach mentioned above to include the effects of modulation, which works well in describing most characteristics of the PDFs that depend on~$\Phi$. However, more research is needed to include the effects of cessations on these PDFs more explicitly, as the current approach has its shortcomings especially in those phases where very high numbers of cessations occur.

We have, furthermore, investigated the reasons for the maximum in $\eta$ at finite $\omega$. We find that, at this $\omega$, a resonance of sorts seems to occur between the ``clean" modulated response of $\theta_d(t)$ (the angular orientation as seen from the rotating frame of reference) and the periods of $\theta_d(t)$ affected by cessations. The sudden azimuthal reorientations of the LSC as a result of cessations of $\delta(t)$ coincidentally synchronize closely in amplitude as well as phase with the clean response, thus reinforcing the modulations instead of interrupting them, as would happen at other $\omega$. This could result in a resonance whereby cessations, instead of interrupting the modulated response, strengthen it continuously, which in its turn increases the probability of further cessation events.

Lastly, we have investigated the effect of modulated rotation rates on heat transfer at lower $Ra$ than for the experiments described above, to explore the potential effects of modulation in absence of an LSC, in a modulation range where the adiabatic response of $Nu$ to changes in the rotation rate is relatively strong. We find that external modulated rotations also result in a modulated response in the Nusselt number $Nu$. Increased modulations turn out to slightly suppress the average $Nu$, with maximum values remaining largely unaffected but minimum values being significantly reduced under increasing modulation rates, as well as to increase its phase delay with respect to $\Omega(t)$. The suppression of heat transfer under modulated rotation is an intriguing phenomenon that in our point of view merits further experimental and numerical research. 

Future research from the authors will work towards extending the ranges of experimental parameters of RB convection in which the dynamical and statistical behaviour of thermal convection and heat transfer can be studied. For example, the current paper discusses the dynamical responses of $\dot{\theta}(t)$ and $\delta(t)$ in a range of rotation rates in which their responses are strong and roughly monotonic with $\Omega(t)$, based on their adiabatic responses, which simplifies the analysis of amplitude and phase responses. However, one could change the range of modulation such that $\dot{\theta}(t)$ and $\delta(t)$ might exhibit a wealth of nonlinear behaviour that goes beyond the current study in complexity. It would be worthwhile to investigate whether the identified mechanisms of phase and amplitude response and statistical behaviour from the current study would still hold in such regimes. We also recommend to more closely align this stream of research with the potential interests from the geophysical and astrophysical communities, by focusing specifically on parameter ranges and/or adapted experimental geometries with higher relevance in geo- and astrophysics.

Furthermore, we believe that the results found on heat transfer suppression are extremely interesting and endeavour to further investigate this subject, as the current study did not include a modeling approach to explain the physics behind this phenomenon. In particular, it would be worth investigating whether the trends found will stand up to scrutiny in other parameter ranges of $Ra$ and $\omega/\Omega_0$. Complementing experimental studies with, for instance, DNS could be highly valuable in identifying the precise physical mechanisms responsible for the trends found in this study.

\section*{Acknowledgements}
This work was supported by the National Science Foundation of China through grant no.\ 11202151, 11572230 and 1561161004.

\section*{Appendix A: Control of modulated rotation}

The rotating table used in our setup could only switch frequencies in a discrete manner, with increments of 0.06$^{\circ}$/s. In order to perform libration, therefore, we needed to approximate a sinusoidal modulation $\Omega(t)$ by a discrete modulation, consisting of a number of discrete frequencies, denoted $F_n = n \times 0.06\pi/180$~rad/s. 

As an example, in our experiments on LSC dynamics and statistics, the librational range was $0.33 < 1/Ro < 0.51$, corresponding (at $\Delta T = 16$~K) to $F_{n,max} = F_{120}$ and $F_{n,min} = F_{78}$. Thus, a sinusoidal modulation for this run was modeled by tuning the frequency down from $F_{120}$ to $F_{78}$ (in 43 steps), then tuning the frequency back to $F_{120}$ (in the same 43 steps in reverse order), and repeating this cycle.

The angle $\Theta_n$ that the table should cover while rotating at each frequency level $F_n$ between $F_{120}$ and $F_{78}$ was determined as follows. 
Assuming the modulation frequency $\omega$ is known, the exact, smooth librational frequency $\Omega(t)$ corresponding to the desired range of $1/Ro$ and the desired~$\omega$ is of course given by $\Omega(t) = \Omega_0[1 + \beta\cos{(\omega t)}]$. 
We start out at the maximum level, $F_{120}$. At this level, we let the table cover half the angle that would be covered at rotation rate $\Omega(t)$ while the latter is larger than $F_{119.5}$ (this ''halfway level'' is imaginary, in the sense that the table could not actually rotate at this value). Thus, 

\begin{equation}
	\Theta_{120} = \int_{0}^{t(\Omega(t) = F_{119.5})} \Omega(t) dt.
\end{equation}
After an angle $\Theta_{120}$ has been covered, we switch to $F_{119}$ and let the table cover the same angle as would be covered at rotation rate $\Omega(t)$ between the values $F_{119.5}$ and $F_{118.5}$:

\begin{equation}
	\Theta_{119} = \int_{t(\Omega(t) = F_{119.5})}^{t(\Omega(t) = F_{118.5})} \Omega(t) dt.
\end{equation}
Subsequently, we switch to $F_{118}$, and let the table cover the same angle as would be covered at rotation rate $\Omega(t)$ between the two frequency levels $F_{118.5}$ and $F_{117.5}$. Et cetera, down to (and including) frequency level $F_{79}$.
Finally, we switch to the lowest level $F_{78}$, and let the table cover half the angle that would be covered at rate $\Omega(t)$ below the value corresponding to frequency level~$F_{78.5}$.

\begin{figure}
	\centering
		\includegraphics[width=1.0\textwidth]{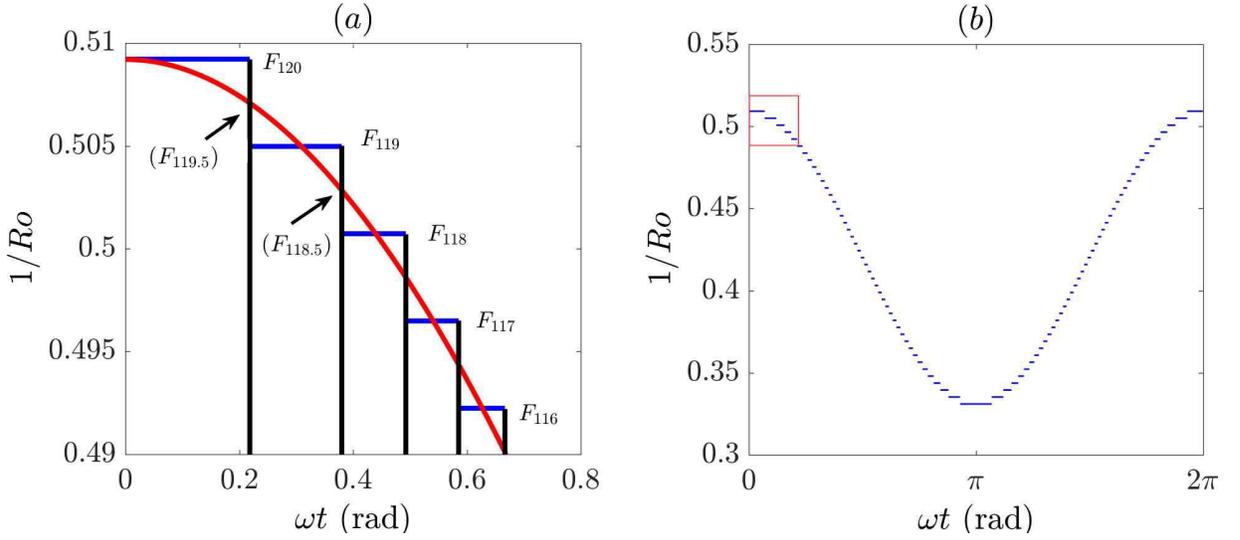}
		\caption{(a) An illustration of the calculation of appropriate timings to model a smooth spin-down with a discrete set of frequencies. The blue lines indicate these discrete frequencies, whereas the red smooth curve is the exact $\Omega(t)$. Sets of ``halfway frequencies'' between two $F$-values are used to calculate the time that $\Omega(t)$ spends between those two halfway levels, and from that, the net angle $\Theta_n$ is calculated that should be covered at each discrete frequency. This, in turn, determines the time spent rotating at each discrete level~$F_n$. (b) The complete librational cycle, with the red box indicating the section shown in (a).}
	\label{fig:Figure_26}
\end{figure}

In Fig.~\ref{fig:Figure_26}(a), we illustrate this procedure by showing the smooth $\Omega(t)$ and its intersections with a number of halfway levels. In this Figure, the angles $\Theta_n$ correspond to the areas underneath the red curve and sandwiched between neighbouring vertical lines.
The spin-up part of the cycle is performed by reversing the spin-down part.

Fig.~\ref{fig:Figure_26}(b) shows the discrete frequencies (expressed as discrete $1/Ro$-values) during an example librational cycle against dimensionless time.
Since our approach is based on setting the angle of rotation~$\Theta_n$, not the time of rotation at each level~$F_n$, the actual period of modulation resulting from this method is slightly different from the theoretical value $2\pi/\omega$ (by less than $\lesssim 0.15\%$ in all cases). However, since the small, but finite switching time between different $F_n$ already introduced an error of at least the same order by itself, this small discrepancy had to be accounted for in any case, i.e.\ for constructing ensemble responses, and the subsequent analysis.
Thus, our method described above, which constructs the stepped libration by requiring that the total angle $\sum_n \Theta_n$ covered during one modulation period is equal to the net theoretical angle $\int_0^{2\pi/\omega} \Omega(t) dt$, was deemed by us to be more useful than trying to keep the modulation period equal to its theoretical value $2\pi/\omega$, when this value would contain an error in any case due to the switching time between discrete frequency levels.

\section*{Appendix B: Calculation of phase responses}

In this Appendix, we explain how we calculate the mean and standard deviation of the phase shifts~$\phi_{\dot{\theta}}$ and~$\phi_{\delta}$.
First, the cross-covariances of $\dot{\theta}(t)$ and $\delta(t)$ with $\Omega(t)$ were computed separately for each event-less period in each experiment. This was done as follows. Each response of $\dot{\theta}(t)$ and $\delta(t)$ inbetween two maxima in $\Omega(t)$ was copied and repeated~$N$ times (we used $N = 10$); subsequently, the cross-covariance between such a repeated period and $N$~periods of $\Omega(t)$ was calculated. The peak therein was identified, and the location of this peak was translated to a phase shift. 
In this way, every period of $\Omega(t)$ that did not contain an event, yielded a value for $\phi_{\dot{\theta}}$ and $\phi_{\delta}$. 
Two additional criteria were applied to filter out noise and unidentified events, respectively:
\begin{enumerate}
	\item We found that the normalized cross-covariance of $\Omega(t)$ with a stochastic vector (consisting of random numbers from a normal distribution) could show peaks with a magnitude of close to 0.1~at the 99\% confidence level. When the magnitude of the peak in a cross-covariance vector was lower than~0.1, therefore, we did not use it.
	\item Secondly, the ensemble behaviour is characterized by a certain phase bandwidth. Individual responses that fall far outside of this band, even if they pass the significance test, do not represent a response to the sinusoidal forcing, but rather a strong deviation caused by a cessation of the LSC that went undetected by the criterion $\delta < \delta_c$.
	We therefore do not use those periods in which the phase shift deviates by more than $\pm \pi/2$ from the modal value of the phase shift. This is based on the empirical observation that the phase bandwidth is less than~$\pi$.
\end{enumerate}
Clearly, these criteria (as well as those mentioned before for taking out events) can overlap, since, for example, a response affected by a cessation can easily show a high deviation in the calculated phase shift. We illustrate the effect of our filtering methods here with a visual example.

\begin{figure}
	\centering
		\includegraphics[width=0.45\textwidth]{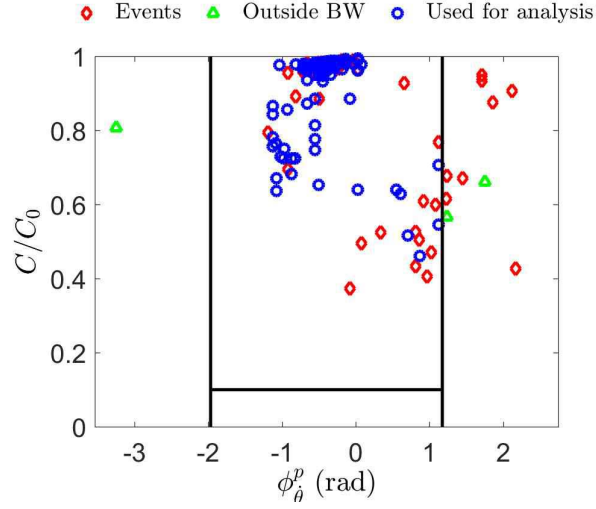}
	\caption{The values of $C/C_0$ of the peak in the cross-covariance of $\dot{\theta}$ with $\Omega(t)$ in the experiment with $\omega/\Omega_0 = 1/8$, plotted against the corresponding value of $\phi_{\dot{\theta}}^p$.}
	\label{fig:Figure_27}
\end{figure}

The most straightforward way to visualize this is by a scatterplot of the normalized value of maximum cross-covariance (denoted $C/C_0$) of either $\dot{\theta}(t)$ or $\delta(t)$ with $\Omega(t)$, against the associated phase shift $\phi_{\dot{\theta}}$ or $\phi_{\delta}$ (calculated from the location of this maximum). We give such a plot in Fig.~\ref{fig:Figure_27} for $\phi_{\dot{\theta}}^p$ (here, the $p$ stands for ``probe", representing the fact that this value has not yet been corrected for the thermal diffusion time between probe and fluid) corresponding to the mid-thermistor data from an experiment with $\omega/\Omega_0 = 1/8$.
In this plot, the data points left out of the analysis by the event-based criteria are plotted as diamonds. The phases of these data points show a lot of scatter, as can be expected. In total, 44 out of 155 periods (28\%) are affected by events and therefore not used in the calculation of the phase and amplitude responses.

It can be seen that there is a large ``point cloud" near $\phi_{\dot{\theta}}^p \approx -0.5$ and with $C/C_0 \approx 1$. These points correspond to responses in $\dot{\theta}$ that are excellently correlated with $\Omega(t)$; they constitute the bulk of the points in this Figure, confirming that there is an ``ensemble phase" of rather limited bandwidth.

The vertical black lines indicate a bandwidth (BW) of $\pi$ centered around the most commonly found value of $\phi_{\dot{\theta}}$, which lies somewhere in the point cloud. Points outside of this bandwidth (and not yet discarded by the event-based criterion), plotted as triangles, are left out of the analysis as well, as they are likely to have little to do with the ensemble response - in fact they are likely to be event-affected, but simply not captured by the (somewhat arbitrary) criteria for identifying events.
In this particular run, this only applies to 3 points out of 111, less than 3\% (not counting the points already discarded from the event-based criterion); however, leaving them out decreases the standard deviation of $\phi_{\dot{\theta}}$ by 25\%.

The horizontal black line indicates the threshold of significance of $C/C_0$; in this particular experiment, every response in $\dot{\theta}$ passed this significance test. The points plotted as circles pass each criterion, and so are used in the calculation of the mean and the standard deviation of $\phi_{\dot{\theta}}^p$. In this case, $\phi_{\dot{\theta}}^p = 0.42 \pm 0.42$, with roughly 30\% of the points being left out of the analysis.

\begin{figure}
	\centering
                   \includegraphics[width=0.8\textwidth]{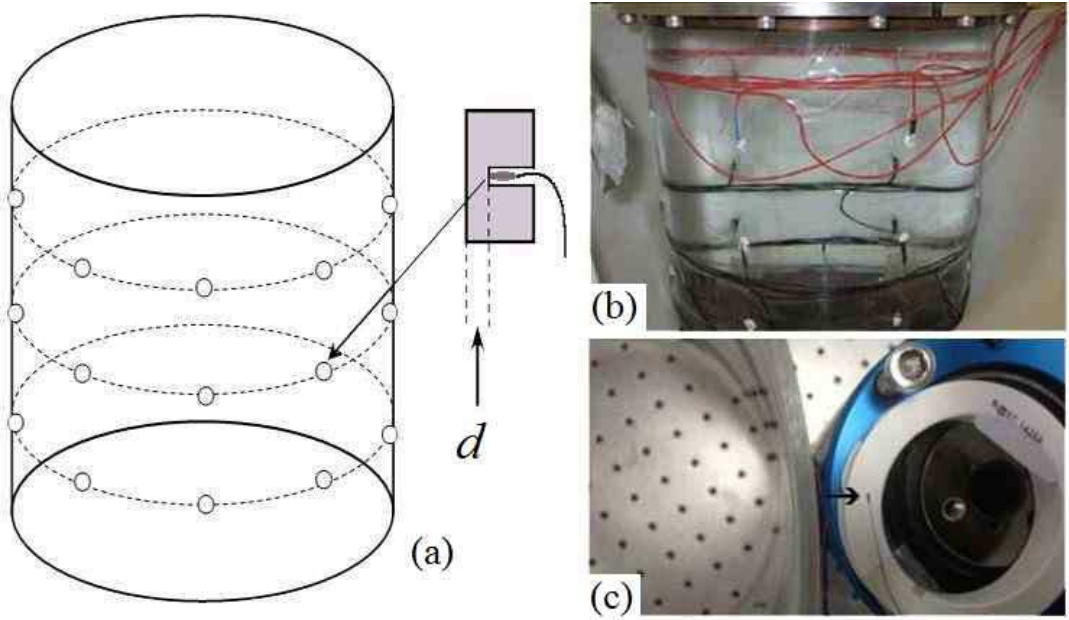}
	\caption{(a) Schematic of the convection cell and positions of the 24~thermistors embedded in the sidewall. (b) Side-view photo of the sidewall with the thermistors installed. (c) Photo of a thermistor (indicated by the black arrow) next to a portion of the sidewall (top view). A stainless steel screw size with 6~mm diameter indicates the scale.}
	\label{fig:Figure_28}
\end{figure}

The effect of thermal diffusion time between thermistor and fluid could account for a slight additional time delay. For proper analysis, this effect needed to be accounted for as well. Here we explain in more detail how this has been done.
There were 24~blind holes in the sample sidewall that had been carefully machined into it from the outside. The ends of these holes had a distance of $d = 0.8 \pm 0.1$~mm from the fluid surface, as indicated in~Fig.~\ref{fig:Figure_28}(a). The sidewall of the sample was made of a cylindrical Lexan plastic tube, with a wall thickness of 4.0~mm and thermal diffusivity $\kappa = 0.144$~mm$^2$/s at 25$^{\circ}$C \cite{BL03}. The temperature probes we used consisted of a thermistor bead (BetaTHERM, type G22K7MCD419) welded to insulated extension leads. The probes had a diameter of 0.38~mm, and the bead was located within 0.2~mm of the end of the probe. The thermal response time of these thermistors was 30~milliseconds. The thermistors were placed into the blind holes in the sample sidewall until they touched the inner ends. To ensure good thermal contact, a thin layer of thermally conductive paste was spread around the surface of the thermistors and filled the leftover empty volume in the holes.

 We estimate the time delay in our temperature measurement to be caused by the finite thermal diffusivity of the sidewall and by the aforementioned response time of the thermistors. Based on the data provided above, we determine that the time delay $\tau_{sw}$ is mainly due to the thermal diffusion time in the sidewall:

\begin{equation}
\tau_{sw} = d^2/\kappa \approx 4.6 \mbox{ s}.
\end{equation}
Thus, we correct our inferred phase responses $\phi_{\dot{\theta}}^p$ and $\phi_{\delta}^p$ as follows to obtain the actual phase shifts $\phi_{\dot{\theta}}$ and $\phi_{\delta}$:

\begin{equation}
\phi_{\dot{\theta}} = \phi_{\dot{\theta}}^p + \omega\tau_{sw} \mbox{ ; } \phi_{\delta} = \phi_{\delta}^p + \omega\tau_{sw}.
\end{equation}
We find that the time lag $\tau_{sw}$ due to the thermal diffusion of the sidewall is smaller than the standard deviation in $\phi_{\dot{\theta}}^p/\omega$ and $\phi_{\delta}^p/\omega$. These results suggest that the time-delay caused by thermal diffusion within the sidewall produces a noticeable phase lag only in a very high regime of $\omega$, and has insignificant effects on the phase response data at low modulation frequencies.

\newpage

%

\end{document}